\documentclass[a4paper,useAMS,usenatbib]{mn2e}

\usepackage{psfig}

\newif\iffigs
\figstrue
\iffigs
\fi

\def\drawing #1 #2 #3 {
\begin{center}
\setlength{\unitlength}{1mm}
\begin{picture}(#1,#2)(0,0)
\put(0,0){\framebox(#1,#2){#3}}
\end{picture}
\end{center} }

\title{Reconstruction of the early Universe as a convex optimization
problem}

\author[Y.~Brenier et al.]%
{Y.~Brenier,$^1$ 
 U.~Frisch,$^{2,3}$\thanks{E-mail: uriel@obs-nice.fr}
 M.~H\'{e}non,$^2$ 
 G.~Loeper,$^1$ 
 S.~Matarrese,$^4$ \newauthor
 R.~Mohayaee,$^2$ 
 A.~Sobolevski\u{\i}$^{2,5}$\\ \\ 
 $^1$ CNRS, UMR 6621, Universit\'e de Nice-Sophia-Antipolis, 
  Parc Valrose, 06108 Nice Cedex 02, France\\ 
 $^2$ CNRS, UMR 6529, Observatoire de la C{\^o}te d'Azur, BP~4229, 
  06304 Nice Cedex 4, France\\ 
 $^3$ Institute for Advanced Study, Einstein Drive, Princeton, 
  NJ 08540, USA\\ 
 $^4$ Dipartimento di Fisica `G. Galilei' and INFN, Sezione di
  Padova, via Marzolo 8, 35131-Padova, Italy\\ 
 $^5$ Department of Physics, M. V. Lomonossov Moscow University, 
  Leninskie Gory, 119992 Moscow, Russia}

\begin{document}

\maketitle

\begin{abstract}
  We show that the deterministic past history of the Universe can be
  uniquely reconstructed from the knowledge of the present mass
  density field, the latter being inferred from the 3D distribution of
  luminous matter, assumed to be tracing the distribution of dark
  matter up to a known bias. Reconstruction ceases to be unique below
  those scales -- a few Mpc -- where multi-streaming becomes
  significant.  Above $6\,h^{-1}$\, Mpc we propose and implement an
  effective Monge--Amp\`ere--Kantorovich method of unique
  reconstruction.  At such scales the Zel'dovich approximation is well
  satisfied and reconstruction becomes an instance of optimal mass
  transportation, a problem which goes back to \citet{M81}.  After
  discretization into $N$ point masses one obtains an assignment
  problem that can be handled by effective algorithms with not more
  than $O(N^3)$ time complexity and reasonable CPU time requirements.
  Testing against $N$-body cosmological simulations gives over 60\% of
  exactly reconstructed points.

  We apply several interrelated tools from optimization theory that
  were not used in cosmological reconstruction before, such as the
  Monge--Amp{\`e}re equation, its relation to the mass transportation
  problem, the Kantorovich duality and the auction algorithm for
  optimal assignment. Self-contained discussion of relevant notions
  and techniques is provided.
\end{abstract}


\begingroup
 
\newcommand   {\J}{{\bmath{J}}}
\newcommand   {\m}{{\bmath{m}}}
\newcommand   {\U}{{\bmath{U}}}
\renewcommand {\u}{{\bmath{u}}}
\renewcommand {\v}{{\bmath{v}}}
\newcommand   {\x}{{\bmath{x}}}
\newcommand   {\y}{{\bmath{y}}}
\newcommand   {\q}{{\bmath{q}}}
\renewcommand {\r}{{\bmath{r}}}
\newcommand   {\s}{{\bmath{s}}}
\renewcommand {\bxi}{{\bmath{\xi}}}

\newcommand{\myinf}{\mathop{\mathrm{inf\vphantom{p}}}}
\renewcommand {\b}{{\bmath{b}}}
\renewcommand {\c}{{\bmath{c}}}

\newcommand   {\phig}{\varphi_{\mathrm{g}}}
\newcommand   {\phiv}{\varphi_{\mathrm{v}}}

\newcommand   {\dtau}{{\partial_\tau}}
\newcommand   {\Dtau}{{D_\tau}}
\newcommand   {\ddtau}{{\partial^2_\tau}}
\newcommand   {\DDtau}{{D^2_\tau}}
\newcommand   {\gradx}{{\nabla_{\x}}}
\newcommand   {\lapx}{{\nabla^2_{\x}}}
\newcommand   {\gradq}{{\nabla_{\q}}}
\newcommand   {\lapq}{{\nabla^2_{\q}}}
\newcommand   {\iint}{\int\!\!\int}

\newcommand{\D}{\mathrm{d}^3}
\renewcommand{\d}{\mathrm{d}}

\newcommand{\eqref}[1]{(\ref{#1})}

\section{Introduction}
\label{s:intro}

Can one follow back in time to initial locations the highly structured
present distribution of mass in the Universe, as mapped by redshift
catalogues of galaxies?  At first this seems an ill-posed problem
since little is known about the peculiar velocities of galaxies, so
that equations governing the dynamics cannot just be integrated back
in time.  In fact, it is precisely one of the goals of reconstruction
to determine the peculiar velocities.  Since the pioneering work of
\citet{P89}, a number of reconstruction techniques have been proposed,
which frequently provided non-unique answers.\footnote{We put the
present work in context of several important existing techniques in
Section~\ref{s:other}.}

Cosmological reconstruction should however take advantage of our knowledge
that the initial mass distribution was quasi-uniform at baryon-photon
decoupling, about 14~billion years ago \citep[see, e.g.,][]{SB94}.  In a
recent Letter to Nature \citep{FMMS02}, four of us have shown that, with
suitable assumptions, this \textit{a~priori} knowledge of the initial density
field makes reconstruction a well-posed instance of what is called the optimal
mass transportation problem.

A well-known fact is that, in an expanding universe with self-gravitating
matter, the initial velocity field is `slaved' to the initial gravitational
field, which is potential; both fields thus depend on a single scalar
function. Hence the number of unknowns matches the number of constraints,
namely the single density function characterising the present distribution of
mass.

This observation alone, of course, does not ensure uniqueness of the
reconstruction.  For this, two restrictions will turn out to be
crucial.  First, from standard redshift catalogues it is impossible to
resolve individual streams of matter with different velocities if they
occupy the same space volume. This `multi-streaming' is typically
confined to relatively small scales of a few megaparsecs~(Mpc), below
which reconstruction is hardly feasible.  Second, to reconstruct a
given finite patch of the present Universe, we need to know its
initial shape at least approximately.

It is our purpose in the present paper to clarify the physical nature
of the factors permitting a unique reconstruction and of obstacles
limiting it, and to give a detailed account of the way some recent
developments in the optimal mass transportation theory are
applicable. (Fig.~\ref{f:deblairemblai} may give the reader some
feeling of what mass transportation is about.)
\begin{figure}
  \iffigs
    \centerline{\psfig{file=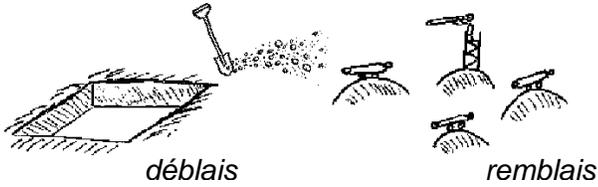,width=8cm}}
  \fi
  \vspace{2mm}
  \caption{A sketch of Monge's mass transportation problem in which
    one searches the optimal way of transporting earth from cuts
    (\textit{d\'eblais}) to fills (\textit{remblais}), each of
    prescribed shape; the cost of transporting a molecule of earth is
    a given function of the distance. The MAK method of reconstructing
    the early Universe described in this paper corresponds to a
    quadratic cost.}
  \label{f:deblairemblai}
\end{figure}

The paper is organized as follows.  In Section~\ref{s:reconstruction}
we formulate the reconstruction problem in an expanding universe and
state the main result about uniqueness of the solution.

In the next three sections we devise and test a reconstruction
technique called MAK (for Monge--Amp\`ere--Kantorovich) within a
restricted framework where the Lagrangian map from initial to present
mass locations is taken potential.  In Section~\ref{s:makcontinuous}
we discuss the validity of the potentiality assumption and its
relation to various approximations used in cosmology; then we derive
the Monge--Amp\`ere equation, a simple consequence of mass
conservation, introduce its modern reformulation as a
Monge--Kantorovich problem of optimal mass transportation and finally
discuss different limitations on uniqueness of the reconstruction.  In
Section~\ref{s:makdiscrete} we show how discretization turns
optimization into an instance of the standard assignment problem; we
then present effective algorithms for its solution, foremost the
`auction' algorithm of D.~Bertsekas.  Section~\ref{s:testing} is
devoted to testing the MAK reconstruction against $N$-body
cosmological simulations.

In Section~\ref{s:selfgravitating}, we show how the general case,
without the potentiality assumption, can also be recast as an
optimization problem with a unique solution and indicate a possible
numerical strategy for such reconstruction. In Section~\ref{s:other}
we compare our reconstruction method with other approaches in the
literature.  In Section~\ref{s:conclusion} we discuss perspectives and
open problems.

A number of topics are left for appendices.  In
Appendix~\ref{a:cosmobasics} we derive the Eulerian and Lagrangian
equations in the form used throughout the paper (and provide some
background for non-cosmologists).  Appendix~\ref{a:historymongeetc} is
devoted to the history of optimal mass transportation theory, a
subject more than two centuries old \citep{M81}, which has undergone
significant progress within the last two decades.
Appendix~\ref{a:duality} is a brief elementary introduction to the
technique of duality in optimization, which we use several times
throughout the paper.  Appendix~\ref{a:details} gives details of the
uniqueness proof that is only outlined in
Section~\ref{s:selfgravitating}.

 Finally, a word about notation (see also
Appendix~\ref{a:cosmobasics}).  We are using comoving coordinates
denoted by~$\x$ in a frame following expansion of the Universe.  Our
time variable is not the cosmic time but the so-called linear growth
factor, here denoted by~$\tau$, whose use gives to certain equations
the same form as for compressible fluid dynamics in a non-expanding
medium.  The subscript~$0$ refers to the present time (redshift
$z=0$), while the quantities evaluated at the initial epoch take the
subscript or superscript `in.'  Following cosmological usage, the
Lagrangian coordinate is denoted $\q$.

\section{Reconstruction in an expanding universe}
\label{s:reconstruction}

The most widely accepted explanation of the large-scale structure seen
in galaxy surveys is that it results from small primordial
fluctuations that grew under gravitational self-interaction of
collisionless cold dark matter (CDM) particles in an expanding
universe \citetext{see, e.g., \citet{BSGS02} and references
therein}. The relevant equations of motion, derived in
Appendix~\ref{a:cosmobasics}, are the Euler--Poisson
equations\footnote{Also often called the Euler equations.} written
here for a flat, matter-dominated Einstein--de~Sitter universe
\citep[for more general case see, e.g.,][]{CLMM95}:
\begin{eqnarray}
  \label{repcomoveuler}
  \dtau\v+(\v\cdot\gradx)\v 
  &=& -\frac{3}{2\tau}(\v+\gradx\phig),\\
  \label{repcomovcontinuity}
  \dtau\rho+\gradx\cdot(\rho\v) &=& 0,\\
  \label{repcomovpoisson}
  \lapx\phig &=& \frac{\rho - 1}{\tau}.
\end{eqnarray}
Here $\v$ denotes the velocity, $\rho$ denotes the density
(normalized by the background density $\bar\varrho$) and $\phig$ is a
rescaled gravitational potential. All quantities are expressed in
comoving spatial coordinates $\x$ and  linear growth factor $\tau$,
which is used as the time variable; in particular, $\v$ is the
Lagrangian $\tau$-time derivative of the comoving coordinate of a
 fluid element.

\subsection{Slaving in early-time dynamics and its fossils}
\label{s:early}

The right-hand sides of the momentum and Poisson equations
\eqref{repcomoveuler} and~\eqref{repcomovpoisson} contain denominators
proportional to $\tau$.  Hence, a necessary condition for the problem not to
be singular as $\tau\to 0$ is
\begin{equation}
  \v_{\rm in}(\x)+\gradx\varphi^{\rm in}_{\rm g}=0,\qquad 
  \rho_{\rm in}(\x) = 1.
  \label{slaving}
\end{equation}
In other words, (i) the initial velocity must be equal to (minus) the
gradient of the initial gravitational potential and (ii) the initial
normalized mass distribution is uniform.  We shall refer to these
conditions as \emph{slaving}. Note that the density contrast $\rho-1$
vanishes initially, but the rescaled gravitational potential and the
velocity, as defined here, stay finite thanks to our choice of the
linear growth factor as time variable. Therefore we refer to the
initial mass distribution as `quasi-uniform.'

In the sequel, when we mention the Euler--Poisson \emph{initial-value
problem}, it is always understood that we start at $\tau =0$ and assume
slaving.  Hence we are extending the Newtonian matter-dominated
post-decoupling description back to $\tau=0.$ By examination of the Lagrangian
equations for $\x(\q,\tau)$ near $\tau=0$, which can be linearized because the
displacement $\x-\q$ is small, it is easily shown that slaving implies the
absence of the `decaying mode,' which behaves as $\tau^{-3/2}$ in an
Einstein--de~Sitter universe and is thus singular at $\tau=0$ (for details see
Appendix~\ref{a:cosmobasics}).

Slaving is also a sufficient condition for the initial problem to be well
posed.  It is indeed easily shown recursively that
\eqref{repcomoveuler}--\eqref{repcomovpoisson} admit
a solution in the form of a formal Taylor series in $\tau$ \citep[a
related expansion involving only potentials may be found
in][]{CLMM95}:
\begin{eqnarray}
  \v(\x,\tau)&=&\v^{(0)}(\x)+\tau\v^{(1)}(\x)+\tau^2\v^{(2)}(\x)+\cdots,
  \label{vexp}\\
  \phig(\x,\tau)&=&\phig^{(0)}(\x)+\tau\phig^{(1)}(\x)+\tau^2\phig^{(2)}(\x)+\cdots,
  \label{phigexp}\\
  \rho(\x,\tau)&=&1+\tau\rho^{(1)}(\x)+\tau^2\rho^{(2)}(\x)+\cdots.
  \label{rhoexp}
\end{eqnarray}
 Furthermore, $\v^{(n)}(\x)$ is easily shown to be curl-free for any~$n$.

Several important consequences of slaving extend to later
times as `fossils' of the earliest dynamics.  First, as already
stressed in the Introduction, the whole dynamics is determined by only
one scalar field (e.g., the initial gravitational potential) which we
can hope to determine from the knowledge of the present density field.

Second, slaving trivially rules out multi-streaming up to the time of
 formation of caustics. Since we are working with collisionless matter,
the dynamics should in principle be governed by the
Vlassov--Poisson\footnote{Actually written for the first time by
\citet{J19}.} kinetic equation which allows at each $(\x,\tau)$ point
a non-trivial distribution function $f(\x,\v,\tau)$.  Slaving selects
a particular class of solutions for which the distribution function is
concentrated on a single-speed manifold, thereby justifying the use of
the Euler--Poisson equation without having to invoke any
hydrodynamical limit \citep[see, e.g.,][]{VDFN94,CLMM95}.

Third, it is easily checked from \eqref{repcomoveuler} that the
initial slaved velocity, which is obviously curl-free, remains so for
all later times (up to formation of caustics). 
Note that this vanishing of the curl holds in Eulerian coordinates. A similar
property in Lagrangian coordinates can only hold approximately but will play
an important role in the sequel (Section~\ref{s:makcontinuous}).

\subsection{Formulation of the reconstruction problem}
\label{s:formulation}

The present Universe is replete with high-density structures: clusters
(point-like objects), filaments (line-like objects) and perhaps sheets
or walls.\footnote{Whether the Great Wall and the Sculptor Wall are
sheet-like or filament-like is a moot point \citep{SSSF98}.}

The internal structure of such \emph{mass concentrations} certainly
displays multi-streaming and cannot be described in terms of a
single-speed solution to the Euler--Poisson 
equations. In $N$-body simulations, multi-stream regions are usually
 found to be of relatively small extension in one or several space
directions, typically not more than a few Mpc, 
and hence have a small volume, although they contain a significant
 fraction of the total mass \citep[see, e.g.][]{WG90}.

In order not to have to deal with tiny multi-stream regions, we
replace the true mass distribution by a `macroscopic' one which has a
regular part and a singular (collapsed) part, the latter concentrated
on objects of dimension less than three, such as points or lines.

The general problem of reconstruction is to find as much information as
possible on the history of the evolution that carries the initial uniform
density into the present macroscopic mass distribution, including the
evolution of the velocities.  In principle we would like to find a solution of
the Euler--Poisson initial-value problem leading to the present density field
$\rho_{0}(\x)$.

A more restricted problem, which we call the `displacement
reconstruction,' is to find the Lagrangian map $\q \mapsto \x(\q)$ and
its inverse $\x\mapsto \q(\x)$, or in other words to answer the
question: where does a given `Monge molecule'\footnote{For Monge and
his contemporaries, the word `molecule' meant a Leibniz infinitesimal
element of mass; see Appendix~\ref{a:historymongeetc}.}  of matter
originate from?  Of course, the inverse Lagrangian map will not be
single-valued on mass concentrations. Furthermore, for practical
cosmological applications, we define a `full reconstruction problem'
as (i) displacement reconstruction and (ii) obtaining the initial and
present peculiar velocity fields, $\v_{\rm in}(\q)$ and $\v_0(\x)$.
 
We shall show in this paper that the displacement reconstruction
problem is uniquely solvable and that the full reconstruction problem
has a unique solution outside of mass concentrations; as to the
latter, they are traced back to \emph{collapsed regions} in the
Lagrangian space whose shape and positions are well defined but the
inner structure of density and velocity fluctuations is irretrievably
lost.

\section{Potential Lagrangian maps: the MAK reconstruction}
\label{s:makcontinuous}

In this and the next two sections we shall assume that the Lagrangian
map from initial positions to present ones is potential
\begin{equation}
  \x = \nabla_{\q}\Phi(\q),
  \label{potential}
\end{equation}
and furthermore that the potential $\Phi(\q)$ is convex, which is, as we shall
see, related to the absence of multi-streaming.

\subsection{Approximations leading to maps with convex potentials}
\label{s:potential}

The motivation for the potential assumption, first used by
\citet{BD89},\footnote{In connection with what was called later the Lagrangian
POTENT method \citep*{DBF90}.} comes from the Zel'dovich approximation
\citep{Z70}, denoted here by ZA, and its refinements. To recall how the ZA
comes about, let us start from the equations for the Lagrangian map
$\x(\q,\tau)$, written in the Lagrangian coordinate $\q$
(Appendix~\ref{a:cosmobasics})
\begin{eqnarray}
  \DDtau\x&=&-\frac{3}{2\tau}(\Dtau\x+\gradx\phig),
  \label{replagnewton}\\
  \lapx\phig&=&\frac{1}{\tau}\left[\left(\det \gradq\x\right)^{-1}-1\right],
  \label{replagpoisson}
\end{eqnarray}
where $\Dtau$ is the Lagrangian time derivative and $\nabla_{x_{i}} \equiv
(\partial{q_{j}}/\partial{x_{i}})\nabla_{q_{j}}$ is the Eulerian gradient
rewritten in Lagrangian coordinates. As shown in Appendix~\ref{a:cosmobasics},
in one space dimension the Hubble drag term $\Dtau\x$ and the gravitational
acceleration term $\gradx\phig$ cancel exactly. Slaving, discussed in
Section~\ref{s:early}, means that the same cancellation holds to leading order
in any dimension for small $\tau$. The ZA extends this as an approximation
without the restriction of small $\tau$.  Within the ZA, the acceleration
$\DDtau\x$ vanishes. 
Hence the Lagrangian map has the form
\begin{eqnarray}
  \x(\q,\tau)&=& \q+\tau (\Dtau\x)_{\rm in}(\q)= 
  \q-\tau\gradq\varphi^{\rm in}_{\rm g}
  (\q)\label{zalagmap1}\\
  \nonumber
  &=&\gradq\Phi(\q,\tau)
\end{eqnarray}
with the potential
\begin{equation}
  \Phi(\q,\tau) \equiv \frac{|\q|^2}{2}-\tau 
  \varphi^{\rm in}_{\rm g}(\q).
\label{zalagmap3}
\end{equation}
 Furthermore, taking the time derivative
of \eqref{zalagmap1}, we see that the velocity $\Dtau\x(\q,\tau)$ is
curl-free with respect to the Lagrangian coordinate $\q$.  

Potentiality of the Lagrangian map (and consequently the Lagrangian
potentiality of the velocity) is perhaps the most important feature of the
ZA. Unlike the vanishing of the acceleration, it does not depend on the choice
of the linear growth factor as the time variable. However,
\emph{unaccelerated} but \emph{vortical} flow would fail to exhibit the
cancellation necessary for the ZA to hold. It is noteworthy that the
potentiality is not limited to the ZA: indeed, the latter can be formulated as
the first order of a systematic Lagrangian perturbation theory in which, up to
second order, the Lagrangian map is still potential under slaving
\citetext{\citealp{MABPR91,B92,BE93}; \citealp*{MSS94,C95}}.

It is well known that the ZA map defined by \eqref{zalagmap1} ceases
in general to be invertible due to the formation of multi-stream
regions bounded by caustics.  Since particles move along straight
lines in the ZA, the formation of caustics proceeds just as in
ordinary optics in a uniform medium in which light rays are also
straight.\footnote{Catastrophe theory has been used to classify the different
types of singularities thus obtained \citep*{AZS82}.}  One of the
problems with the ZA is that caustics, which start as localized
objects, quickly grow in size and give unrealistically large
multi-stream regions. 

A modification of the ZA that has no multi-streaming at all, but sharp mass
concentrations in the form of shocks and other singularities, has been
introduced by \citeauthor{GS84} \citetext{\citeyear{GS84}; \citealp[see
also][]{GSS89,SZ89}}. It is known as the \emph{adhesion model}.  In Eulerian
coordinates it amounts to using a multidimensional Burgers equation
\citep[see, e.g.,][]{FB02}
\begin{equation}
  \dtau\v+(\v\cdot\gradx)\v 
   = \nu \lapx \v, \qquad \v=-\gradx\phiv,
  \label{3dburgers}
\end{equation}
taken in the limit where the viscosity $\nu$ tends to zero.  In
Lagrangian coordinates, the adhesion model is obtained from the ZA by
replacing the velocity potential $\Phi(\q,t)$ given by
\eqref{zalagmap3} by its \emph{convex hull} $\Phi_c(\q,t)$ in the $\q$
variable \citep{VDFN94}.

Convexity is a concept which plays an important role in this paper,
and a few words on it are in order here (see also Appendix~\ref{a:convexity}).
A body in the three-dimensional space is said to be \emph{convex} if, whenever
it contains two points, it contains also the whole segment joining them.  A
function $f(\q)$ is said to be convex if the set of all points lying above its
graph is convex.  The convex hull of the function $\Phi(\q)$ is defined as the
largest convex function whose graph lies below that of $\Phi(\q)$.  In two
dimensions it can be visualized by wrapping the graph of $\Phi(\q)$ tightly
from below with an elastic sheet.

Note that $\Phi(\q,\tau)$ given by \eqref{zalagmap3} is obviously
convex for small enough $\tau$ since it is then very close to the
parabolic function $|\q|^2/2$. 
After caustics form, convexity is lost in the ZA but recovered with
the adhesion model.  It may then be shown that those regions in the
Lagrangian space where $\Phi(\q,t)$ does not coincide with its convex
hull will be mapped in the Eulerian space to sheets, lines and points,
each of which contains a finite amount of mass. At these locations the
Lagrangian map does not have a uniquely defined Lagrangian antecedent
but such points form a set of vanishing volume. 
Everywhere else, there is a unique antecedent and hence no
multi-streaming.

Although the adhesion model has a number of known shortcomings, such
as non-conservation of momentum in more than one dimension, it has
been found to be in better agreement with $N$-body simulations than
the ZA \citep{WG90}. Other single-speed approximations to multi-stream
flow, overcoming difficulties of the adhesion model, are 
discussed e.g.\ by \citet{SS96,BD98,FA02}.
In such models, multi-streaming is completely
suppressed by a mechanism of momentum exchange between neighbouring
streams with different velocities.  This is of course a common
phenomenon in ordinary fluids, where it is due to viscous diffusion;
dark matter is however essentially collisionless and the usual
mechanism for generating viscosity does not operate, so that a
non-collisional mechanism must be invoked.  A qualitative explanation
using the modification of the gravitational forces after the formation
of caustics has been proposed by \citet{SZ89}.  In our opinion the
mechanism limiting multi-streaming to rather narrow regions is poorly
understood and deserves considerable further investigation.

\subsection{The Monge--Amp{\`e}re equation: a consequence of mass
  conservation and potentiality}
\label{s:mae}

We now show that the assumption that the Lagrangian map is derived
 from a convex potential leads to a pair of non-linear partial
differential equations, one for this potential and another for its
Legendre transform.

Let us first assume that the present distribution of mass has no singular
part, an assumption which we shall relax later. Since in our notation the
initial quasi-uniform mass distribution has unit density, mass conservation
implies $\rho_0(\x)\, \D \x = \D \q$, which can be rewritten in terms of the
Jacobian matrix 
$\gradq\x$ as
\begin{equation}
  \label{constraint}
  \det\gradq\x = \frac{1}{\rho_0(\x(\q))}.
\end{equation}
Under the potential assumption \eqref{potential}, this takes the form
\begin{equation}
  \label{MAdir}
  \det (\nabla_{q_i}\nabla_{q_j} \Phi(\q)) =
  \frac1{\rho_0\bigl(\gradq\Phi(\q)\bigr)}.
\end{equation}
A similar equation follows also from Eqs.~(1) and~(2) of \citet{BD89}.

A simpler equation, in which the unknown appears only in the left-hand
side, 
viz Eq.~\eqref{MA} below, is obtained 
 for the potential of the \emph{inverse Lagrangian map} $\q(\x)$. Key is the
observation that the inverse of a map with a convex potential has also a
convex potential,
and that the two potentials are Legendre transforms of each
other.%
\footnote{Besides our problem, this fact prominently appears in two
other fields of physics: in classical mechanics, the Lagrangian and
Hamiltonian functions are Legendre transforms of each other -- their
gradients relate the generalized velocity and momentum -- and so are,
in thermodynamics, the internal energy and the Gibbs potential,
implying the same relation between extensive and intensive parameters
of state.} A purely local proof of this statement is to observe that
potentiality of $\q(\x)$ is equivalent to the symmetry of the
\emph{inverse Jacobian matrix} $\gradx\q$ which follows because it is
the inverse of the symmetrical matrix $\gradq\x$; convexity is
equivalent to the positive-definiteness of these matrices. Obviously
the function
\begin{equation}
  \Theta(\x)\equiv \x\cdot\q(\x) -\Phi(\q(\x)),
  \label{deftheta}
\end{equation}
which is the Legendre transform of $\Phi(\q)$, is the potential for
the inverse Lagrangian map. The modern definition of the Legendre
transformation (see Appendix~\ref{a:convexity}), needed for
generalization to non-smooth mass distributions, is
\begin{eqnarray}
  \Theta(\x)& =& \max_\q\, \x\cdot\q -\Phi(\q),
  \label{legendre1}\\
  \Phi(\q) &= &\max_\x\, \x\cdot\q -\Theta(\x).
  \label{legendre2}
\end{eqnarray}

In terms of the potential $\Theta$, mass conservation is immediately written
as
\begin{equation}
  \label{MA}
  \det(\nabla_{x_i}\nabla_{x_j}\Theta(\x)) = \rho_0(\x).
\end{equation}
This equation, which has the determinant of the second derivatives of
the unknown in the left-hand side and a prescribed (positive) function
in the right-hand side, is called the (elliptic) Monge--Amp\`ere
equation (see Appendix~\ref{a:historymongeetc} for a historical
perspective).

Notice that our Monge--Amp\`ere equation may be viewed as a non-linear
generalization of the Poisson equation (used for reconstruction by
\citet{ND92}; see also Section~\ref{s:perturbative}), to which it
reduces if particles have moved very little from their initial
positions.

In actual reconstructions we have to deal with mass concentration in the
present distribution of matter. Thus the density in the right-hand side of
\eqref{MA} has a singular component (a Dirac distribution concentrated on sets
carrying the concentrated mass) and the potential $\Theta$ ceases to be
smooth. As we now show, a generalized meaning can nevertheless be given to the
Monge--Amp\`ere equation by using the key ingredient in its derivation, namely
mass conservation, in integrated form.

 For a nonsmooth convex potential $\Theta$, taking the gradient
$\gradx\Theta(\x)$ still makes sense if one allows it to be
multivalued at points where the potential is not differentiable. The
gradient at such a point~$\x$ is then the set of all possible slopes
of planes touching the graph of $\Theta$ at~$(\x, \Theta(\x))$ (this
idea is given a precise mathematical formulation in
Appendix~\ref{a:convexity}).
As $\x$ varies over an arbitrary domain $\mathcal{D}_\mathrm{E}$ in the
Eulerian space, its image $\q(\x)$ sweeps a domain
$\q(\mathcal{D}_\mathrm{E})$ in the Lagrangian space, and mass
conservation requires that
\begin{equation}
  \label{MAweak}
  \int_{\mathcal{D}_\mathrm{E}} \rho_0(\x)\, \D\x =
  \int_{\gradx\Theta(\mathcal{D}_\mathrm{E})} \D\q,
\end{equation}
where we take into account that $\q(\x) =
\gradx\Theta(\x)$. Eq.~\eqref{MAweak} must hold for any Eulerian
domain $\mathcal{D}_\mathrm{E}$; this requirement is known as the
\emph{weak formulation} of the Monge--Amp\`ere equation \eqref{MA}. A
symmetric formulation may be written for \eqref{MAdir} in terms
of~$\x(\q) = \gradq\Phi(\q)$. For further material on the weak
 formulation see, e.g., \citet{P78}.

Considerable literature has been devoted to the Monge--Amp\`ere
equation in recent years \citep[see, e.g.,][]{C99,MAE99}. We mention
now a few results which are of direct relevance for the reconstruction
problem.

In a nutshell, one can prove that when the domains occupied by the
mass initially and at present are bounded and convex, the
Monge--Amp\`ere equation -- in its weak formulation -- is guaranteed
to have a unique solution, 
which is smooth unless one or both of the
mass distributions is non-smooth.  The actual construction of this
solution can be done 
by a variational method discussed in the next section.

A similar result holds also when the present density field is periodic
and the same periodicity is assumed for the map.  

Also relevant, as we
shall see in Section~\ref{s:redholes}, is a recent result of
\citet{CL01}: if the Monge--Amp\`ere equation is considered in the
whole space, but the present density contrast $\delta =\rho-1$
vanishes outside of a bounded set, then the solution $\Theta(\x)$ is
determined uniquely up to prescription of its asymptotic behaviour at
infinity, which is specified by a quadratic function of the form
\begin{equation}
  \label{xaxbxc}
  \theta(\x) \equiv \langle \x, A\x\rangle + \langle\bmath{b}, \x\rangle + c, 
\end{equation}
 for some positive definite symmetric matrix $A$ with unit determinant,
vector $\bmath{b}$ and constant $c$.

\subsection{Optimal mass transportation}
\label{s:least}

As we are going to see now, the Monge--Amp\`ere equation \eqref{MA} is
equivalent to an instance of what is called the `optimal mass transportation
problem.'  Suppose we are given two distributions $\rho_{\rm in}(\q)$ and
$\rho_0(\x)$ of the same amount of mass in two three-dimensional convex
bounded domains ${\cal D}_{\rm in}$ and ${\cal D}_{0}$. The optimal mass
transportation problem is then to find the most cost-effective way of
rearranging by a suitable map one distribution into the other, the cost of
transporting a unit of mass from a position~$\q\in{\cal D}_{\rm in}$ to 
$\x\in{\cal D}_{0}$ being a prescribed function $c(\q,\x)$.

Denoting the map by $\x(\q)$ and its inverse $\q(\x)$, 
we can write the problem as the requirement that the cost
\begin{equation}
  I \equiv 
  \int_{{\cal D}_{\rm in}}\!\!\! c(\q,\x(\q))\rho_{\rm in}(\q)\,\D\q
  =
  \int_{{\cal D}_{0}}\!\!\! c(\q(\x),\x)\rho_0(\x)\,\D\x
  \label{optcost}
\end{equation}
be minimum, with the constraints of prescribed `terminal' densities
$\rho_{\rm in}$ and $\rho_0$ and of mass conservation $\rho_{\rm in}(\q)\,
\D\q =\rho_0(\x)\,\D\x$.\footnote{Note that $\x(\q) = \q$ does not solve the
above problem as it violates the latter constraint unless the terminal
densities are identical.}

This problem goes back to \citet{M81} who considered the case of a
linear cost function $c(\q,\x) = |\x-\q|$ (see
Appendix~\ref{a:historymongeetc} and Fig.~\ref{f:deblairemblai}).

 For our purposes, the central result is that \textit{the problem of
finding a potential Lagrangian map with presecribed initial and
present mass density fields is equivalent to a mass transportation
problem with quadratic cost.}  Indeed, it is known \citep{B87,B91}
that, when the cost is a quadratic function of the distance, so that
\begin{equation}
  I=\!
  \int_{{\cal D}_{\rm in}}\!\!\!\!\frac{|\x(\q)-\q|^2\!\!\!}{2}\,
  \rho_{\rm in}(\q)\,\D\q =\!
  \int_{{\cal D}_{0}}\!\!\!\! \frac{|\x-\q(\x)|^2\!\!\!}{2}\,
  \rho_0(\x)\,\D\x,
  \label{quadraticcost}
\end{equation}
the solution $\q(\x)$ to the optimal mass transportation problem is
the gradient of a convex function, which then must satisfy the
Monge--Amp\`ere equation \eqref{MA} by mass conservation.

A particularly simple variational proof can be given for the smooth
case, when the two mutually inverse maps $\x(\q)$ and $\q(\x)$ are
both well defined.

Performing a variation of the map $\x(\q)$, we cause a mass element in
the Eulerian space that was located at $\x(\q)$ to move to $\x(\q) +
\delta\x(\q)$. This variation is constrained not to change the density
 field $\rho_0$. To express this constraint it is convenient to rewrite
the displacement in Eulerian coordinate $\delta\x_{\rm E}(\x) \equiv
\delta\x(\q(\x))$. Noting that the point $\x$ gets displaced into $\y
= \x + \delta\x$, we thus require that $\rho_0(\x)\,\D\x =
\rho_0(\y)\,\D\y$ or
\begin{equation}
  \label{derivmasscons}
  \rho_0(\x) = \rho_0(\x + \delta\x_\mathrm{E}(\x))\, 
  \det\bigl(\gradx(\x + \delta\x_\mathrm{E}(\x))\bigr).
\end{equation}
Expanding this equation, we find that, to the leading order,
\begin{equation}
  \gradx\cdot \left(\rho_0(\x)\, \delta\x_{\rm E}(\x)\right) =0,
  \label{eulerconstraint}
\end{equation}
an equation which just expresses the physically obvious fact that the
mass flux $\rho_0(\x)\, \delta\x_{\rm E}(\x)$ should have zero
divergence.  Performing the variation on the functional $I$ given by
\eqref{quadraticcost}, we get
\begin{eqnarray}
  \nonumber
  \delta I &=& \int_{\mathcal{D}_\mathrm{in}} (\x(\q)-\q) \cdot \delta\x(\q)\,
  \rho_\mathrm{in}(\q)\, \D\q\\
  \label{variation}
  &=& \int_{\mathcal{D}_0} (\x - \q(\x)) \cdot 
  \bigl(\rho_0(\x)\, \delta\x_\mathrm{E}(\x)\bigr)\, \D\x = 0,
\end{eqnarray}
which has to hold under the constraint \eqref{eulerconstraint}. In
other words, the displacement $\x - \q(\x)$ has to be orthogonal (in
the $L_2$ functional sense) to all divergence-less vector fields and,
thus, must be a gradient. Since $\x$ is obviously a gradient, it
 follows that $\q(\x)=\gradx\Theta(\x)$ for a suitable potential
$\Theta$.

It remains to prove the convexity of $\Theta$. First we prove that the
map $\x \mapsto \q(\x) = \gradx\Theta(\x)$ is \emph{monotone}, i.e.,
by definition, that for any $\x_1$ and $\x_2$
\begin{equation}
  \label{monot}
  (\x_2 - \x_1) \cdot (\q(\x_2) - \q(\x_1)) \ge 0.
\end{equation}
Indeed, should this inequality be violated for some
$\overline{\x}_1,\overline{\x}_2$, the continuity of $\q(\x)$ would
imply that 
for all $\x_1, \x_2$ close enough to $\overline{\x}_1, \overline{\x}_2$
\begin{equation}
  \begin{array}{l}
    \displaystyle |\q(\x_1) - \x_1|^2 + |\q(\x_2) - \x_2|^2\\[1.5ex]
    \displaystyle\quad > |\q(\x_2) - \x_1|^2 + |\q(\x_1) - \x_2|^2.
  \end{array}
  \label{utterly}
\end{equation}
This in turn means that if we interchange the
destinations of small patches around $\overline{\x}_1$ and
$\overline{\x}_2$, sending them not to the corresponding patches
around $\q(\overline{\x}_1)$ and $\q(\overline{\x}_2)$ but vice versa,
then the value of the functional $I$ will decrease by a small yet
positive quantity, and therefore it cannot be minimum for the original
map.\footnote{As we shall see in Section~\ref{s:assignment}, the
converse is not true: monotonicity alone does not imply that the
integral $I$ is a minimum; the minimizing map must also be potential.}

To complete the argument, observe that convexity of a smooth function
$\Theta(\x)$ follows if the matrix of its second derivatives
$\nabla_{x_i}\nabla_{x_j}\Theta(\x)$ is positive definite for all~$\x$.
Substituting $\q(\x) = \gradx\Theta(\x)$ into \eqref{monot}, assuming
that $\x_2$ is close to $\x_1$ and Taylor expanding, we find that
\begin{equation}
  (\x_2 - \x_1) \cdot (\nabla_{x_i}\nabla_{x_j} \Theta(\x_1)\, (\x_2 - \x_1))
  \ge 0.
\end{equation}
As $\x_2$ is arbitrary, this  
proves the desired positive definiteness and thus establishes the
equivalence of the Monge--Amp\`ere equation \eqref{MA}
and of the mass transportation problem with quadratic cost.

This equivalence is actually proved under much weaker conditions, not
requiring any smoothness \citep{B87,B91}.  
The proof makes use of the `relaxed' reformulation of the mass
transportation problem due to \citet{K42}.  Instead of solving the
highly non-linear problem of finding a map $\q(\x)$ minimizing the
cost \eqref{optcost} with prescribed terminal densities, Kantorovich
considered the \emph{linear programming} problem of minimizing
\begin{equation}
  \tilde I \equiv \int_{\mathcal{D}_\mathrm{in}}\!\!\int_{\mathcal{D}_0}
  c(\q, \x)\, \rho(\q, \x)\, \D\q\, \D\x,
  \label{optcostrelaxed}
\end{equation}
under the constraint that the joint distribution $\rho(\q, \x)$ is
nonnegative and has marginals $\rho_\mathrm{in}(\q)$ and $\rho_0(\x)$,
the latter being equivalent to
\begin{equation}
  \int_{\mathcal{D}_0}\rho(\q, \x)\,\D\x = \rho_\mathrm{in}(\q),\quad
  \int_{\mathcal{D}_\mathrm{in}}\rho(\q, \x)\,\D\q = \rho_0(\x).
\label{marginals}
\end{equation}
Note that if we assume any of the two following forms for the joint
distribution 
\begin{equation}
  \begin{array}{l}
    \rho(\q,\x)=\rho_0(\x)\,\delta\bigl(\q-\q(\x)\bigr)\;\\[1ex]
    \rho(\q,\x)=\rho_{\rm in}(\q)\,\delta\bigl(\x-\x(\q)\bigr),
  \end{array}
\end{equation}
we find that $\tilde I$ reduces to the cost $I$ as defined in
\eqref{optcost}. This relaxed formulation allowed Kantorovich to
establish the existence of a mimimizing joint distribution.

The relaxed formulation can be used to show that the minimizing
solution actually defines a map, which need not be smooth if one or
both of the terminal distribution have a singular component (in our
case, when mass concentrations are present).  The derivation
\citep{B87,B91} makes use of the technique of duality
(Appendix~\ref{a:dualityproper}), which will also 
appear in discussing algorithms (Section~\ref{s:nuts}) and
reconstruction beyond the potential hypothesis
(Section~\ref{s:selfgravitating}).

We have thus shown that the Monge--Kantorovich optimal mass
transportation problem can be applied to solving the Monge--Amp\`ere
equation.  The actual implementation (Section~\ref{s:makdiscrete}),
done for a suitable discretization, will be henceforth called
Monge--Amp\`ere--Kantorovich (MAK).

\subsection{Sources of uncertainty in reconstruction}
\label{s:redholes}

In this section we discuss various sources of non-uniqueness of the
MAK reconstruction: multi-streaming, collapsed regions, reconstruction
 from a finite patch of the Universe.

We have stated before that our uniqueness result applies only in so
 far as we can treat present-epoch high-density multi-stream regions as
if they were truly collapsed, ignoring their width.  We now give a
simple one-dimensional example of non-uniqueness in which a thick
region of multi-streaming is present.  Fig.~\ref{f:nonuniqumulti}
shows a multi-stream Lagrangian map $x(q)$ and the associated density
distribution; the inverse map $q(x)$ is clearly multi-valued. The same
density distribution may however be generated by a spurious
single-stream Lagrangian map shown on the same figure. There is no way
to distinguish between the two inverse Lagrangian maps if the various
streams cannot be disentangled.
\begin{figure}
  \iffigs
    \centerline{\psfig{file=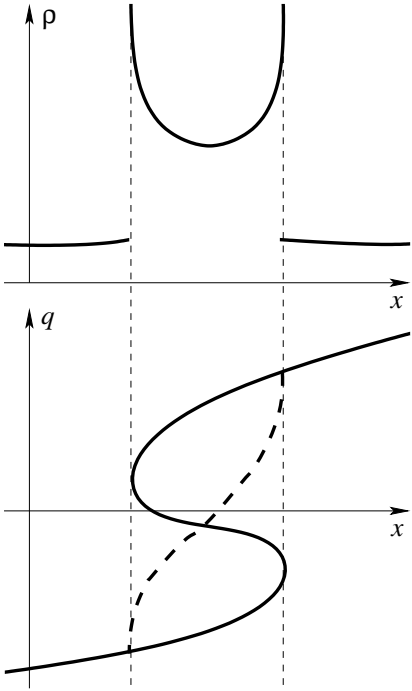,width=5.5cm}}
  \fi
  \vspace{2mm}
  \caption{A one-dimensional example of non-unique reconstruction of
    the Lagrangian map in the presence of multi-streaming. The density
    distribution (upper graph) is generated by a multi-streaming
    Lagrangian map (thick line of lower graph) but may also be
    generated by a spurious single-stream Lagrangian map 
    (dashed line).}
  \label{f:nonuniqumulti}
\end{figure}

Suppose now that the present density has a singular part, i.e.\ there
are mass concentrations present which have vanishing (Eulerian)
volumes but possess finite masses. Obviously any such object
originates from a domain in the Lagrangian space which occupies a
 finite volume.  A one-dimensional example is again
helpful. Fig.~\ref{f:graph} shows a Lagrangian map in which a whole
Lagrangian shock interval $[q_1,q_2]$ has collapsed into a single point
of the $x$~axis. Outside of this point the Lagrangian map is uniquely
invertible but the point itself has many antecedents. Note that the
graph of the Lagrangian map may be inverted by just interchanging the
$q$ and $x$ axes, but its inverse contains a piece of vertical
line. The position of the Lagrangian shock interval which has
collapsed by the present epoch is uniquely defined by the present mass
 field but the initial velocity fluctuations in this interval cannot be
uniquely reconstructed. In particular there is no way to know if
collapse has started before the present epoch. We can of course
arbitrarily assume that collapse has just happened at the present
epoch; if we also suppose that particles have travelled with a
constant speed, i.e.\ use the Zel'dovich/adhesion approximation, then
the initial velocity profile within the Lagrangian shock interval will
be linear (Fig.~\ref{f:graph}). Any other smooth velocity profile
joining the same end points would have points where its slope
(velocity gradient) is more negative than that of the linear profile
(Fig.~\ref{f:graph}) and thus would have started collapse before the
present epoch (in one dimension caustics 
appear at the time which is minus the inverse of the most negative
initial velocity gradient).
\begin{figure}
  \iffigs
    \centerline{\psfig{file=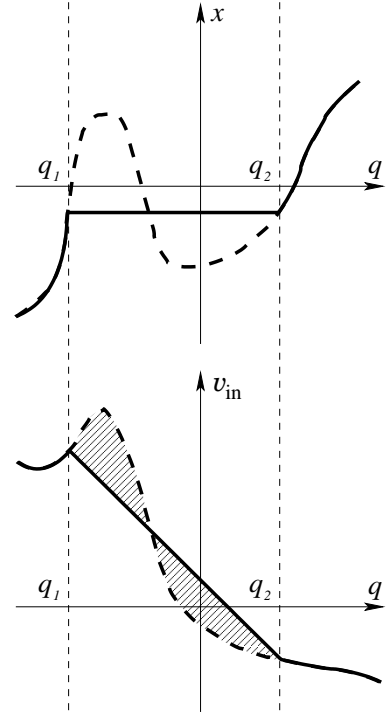,width=5cm}}
  \fi
  \vspace{2mm}
  \caption{Two initial velocity profiles $v_\mathrm{in}(q)$ (bottom,
    solid and dashed lines) leading to the same Lagrangian map $x = q
    + \tau v_\mathrm{in}(q)$ (top, solid line) in the adhesion
    approximation. The Zel'dovich approximation would give
    multistreaming (top, dashed line). Hatched areas (bottom) are
    equal in the adhesion dynamics.}
  \label{f:graph}
\end{figure}

All this carries over to more than one dimension. The MAK
reconstruction gives a unique antecedent for any Eulerian position
outside mass concentrations. Each mass concentration in the Eulerian
space, taken globally, has a uniquely defined Lagrangian antecedent
region but the initial 
velocity field inside the latter is unknown. In other words,
displacement reconstruction is well defined but full reconstruction,
based on the Zel'dovich/adhesion approximation for velocities, is
possible only outside of mass concentrations (note however that
velocities in the Eulerian space are still reconstructed at almost all
points).  We call the corresponding initial Lagrangian domains
\emph{collapsed regions}.

 Finally, we consider a uniqueness problem arising from knowing the
present mass distribution only truncated over a finite Eulerian domain ${\cal
D}_0$, as is necessarily the case when working with a real
catalogue. If we also know the corresponding Lagrangian domain ${\cal
D}_\mathrm{in}$ and both domains are bounded and convex, then
uniqueness is guaranteed (see Section~\ref{s:mae}). What we know for
sure about ${\cal D}_\mathrm{in}$ is its volume, which (in our units)
is equal to the total mass contained in ${\cal D}_0$. Its shape and
position may however be constrained by further information.  For
example, if we know that the typical displacement of mass elements
since decoupling is about ten Mpc in comoving coordinates (see
Section~\ref{s:testing}) and our data extend over a patch of typical
size one hundred Mpc, then there is not more than a ten percent
uncertainty on the shape of ${\cal D}_\mathrm{in}$.  Additional
information about peculiar velocities may also be used to constrain
${\cal D}_\mathrm{in}$.

Note also that a finite-size patch ${\cal D}_0$ with unknown
antecedent ${\cal D}_\mathrm{in}$ will give rise to a unique
reconstruction (up to a translation) if we assume that it is
surrounded by a uniform background extending to infinity.  This is a
consequence of the result of \citeauthor{CL01} mentioned at the end of
Section~\ref{s:mae}. The arbitrary linear term in \eqref{xaxbxc}
corresponds to a translation; as to the quadratic term, it is
constrained by the cosmological principle of isotropy to be exactly
$|\q|^2/2$.

\section{The MAK method: discretization and algorithmics}
\label{s:makdiscrete}

In this section we show how to compute the solution to the
Monge--Amp\`ere--Kantorovich (MAK) problem the known present density
field. First the problem is discretized into an assignment problem
(Section~\ref{s:assignment}), then we present some general tools which make
the assignment problem computationally tractable (Section~\ref{s:nuts}) and
finally we present, to the best of our knowledge, the most
effective method for solving our particular assignment problem, based
on the auction algorithm of D.~Bertsekas (Section~\ref{s:auction}),
and details of its implementation for the MAK reconstruction
(Section~\ref{s:auctionmak}).

\subsection{Reduction to an  assignment problem}
\label{s:assignment}

Perhaps the most natural way of discretizing a spatial mass distribution is to
approximate it by a finite system of identical Dirac point masses, with
possibly more than one mass at a given location. This is compatible both with
$N$-body simulations and with the intrinsically discrete nature of observed
luminous matter. Assuming that we have $N$ unit masses both in the Lagrangian
and the Eulerian space, we may write
\begin{equation}
  \label{discrdistrib}
  \rho_0(\x) = \sum_{i = 1}^N \delta(\x - \x_i),\qquad
  \rho_\mathrm{in}(\q) = \sum_{j = 1}^N \delta(\q - \q_j).
\end{equation}
 For discrete densities of this form, the mass conservation
constraint in the optimal mass transportation problem
(Section~\ref{s:least}) requires that the map $\q(\x)$ induce a
one-to-one pairing between positions of the unit masses in the $\x$
and~$\q$ spaces, which may be written as a permutation of indices that
sends $\x_i$ to~$\q_{j(i)}$. Substituting this into the quadratic cost
 functional \eqref{quadraticcost}, we get
\begin{equation}
  I = \sum_{i = 1}^N \frac{|\x_i - \q_{j(i)}|^2}{2}.
  \label{discraction}
\end{equation}
We thus reduced the problem to the purely
combinatorial one of finding a permutation $j(i)$ (or its inverse
$i(j)$) that minimizes the quadratic cost
 function~\eqref{discraction}.

This problem is an instance of the general \emph{assignment problem}
in combinatorial optimization: for a cost matrix $c_{ij}$, find a
permutation $j(i)$ that minimizes the cost function 
\begin{equation}
  \label{gendiscraction}
  I = \sum_{i = 1}^N c_{i\, j(i)}.
\end{equation}
As we shall see in the next sections, there exist
effective algorithms for finding minimizing permutations.

Before proceeding with the assignment problem, we should mention an
alternative approach in which discretization is performed only in the Eulerian
space and the initial mass distribution is kept continuous and
uniform. Minimization of the quadratic cost function will then give rise to a
tesselation of the Lagrangian space into polyhedric regions which end up
collapsed into the discrete Eulerian Dirac masses. Basically, the reason why
these regions are polyhedra is that the convex potential $\Phi(\q)$ of the
Lagrangian map has a gradient which takes only finitely many values. This
problem, which has been studied by Aleksandrov and Pogorelov \citep[see,
e.g.,][]{P78}, is closely related to Minkowski's \citeyearpar{M97} famous
problem of constructing a convex polyhedron with prescribed areas and
orientations of its faces (in our setting, areas and orientations correspond
to masses and values of the gradient). Uniqueness in the Minkowski problem is
guaranteed up to a translation. Starting with Minkowski's own very elegant
solution, various methods of constructing solutions to such geometrical
questions have been devised. So far, we have not been able to make use of such
ideas in a way truly competitive with discretization in both spaces and
solving then the assignment problem.

The solution to our assignment problem (with quadratic cost) has the
important property that it is monotone: for any two Lagrangian
positions $\q_1$ and $\q_2$, the corresponding Eulerian positions
$\x_1$ and $\x_2$ are such that
\begin{equation}
  (\x_1 - \x_2)\cdot(\q_1 - \q_2) \ge 0.
  \label{discrmonotone}
\end{equation}
This is of course the discrete counterpart of \eqref{monot}. In one dimension,
when all the Dirac masses are on the same line, monotonicity implies that the
leftmost Lagrangian position goes to the leftmost Eulerian position, the
second leftmost Lagrangian position to the second leftmost Eulerian position,
etc. It is easily checked that this correspondence minimizes the cost
\eqref{discraction}.

In more than one dimension, a correspondence between Lagrangian and Eulerian
positions that is just monotone will usually not minimize the cost (a simple
two-dimensional counterexample is given in Fig.~\ref{f:cgniet}).\footnote{Note
that in one dimension, in the continuous case, any map is a gradient and we
have already observed in Section~\ref{s:least} that if a gradient map is
monotone it is the gradient of a convex function.} 
\begin{figure}
  \iffigs
    \centerline{\psfig{file=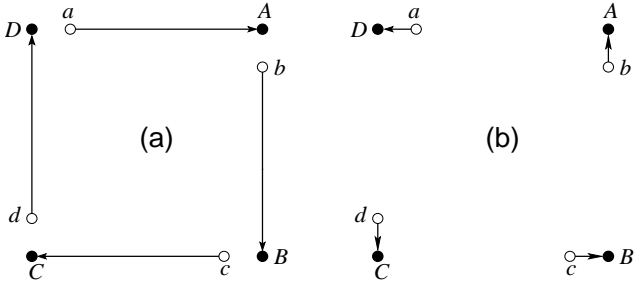,width=8.4cm}}
  \fi
  \vspace{2mm}
  \caption{Two monotone assignments sending white points to black
    ones: (a) an assignment that is vastly non-optimal in terms of
    quadratic cost but cannot be improved by any pair interchange; (b)
    the optimal assignment, shown for comparison.}
  \label{f:cgniet}
\end{figure}
Actually, a much stronger condition, called \emph{cyclic monotonicity}, is
needed in order to minimize the cost. It requires $k$-monotonicity for any $k$
between $2$ and $N$; the latter is defined by taking any $k$ Eulerian
positions with their corresponding Lagrangian antecedents and requiring that
the cost \eqref{discraction} should not decrease under an arbitrary
reassignment of the Lagrangian positions within the set of Eulerian positions
taken. Note that the usual monotonicity corresponds to 2-monotonicity
(stability with respect to pair exchanges).

A strategy called PIZA (Path Interchange Zel'dovich Approximation) for
constructing monotone correspondences between Lagrangian and Eulerian
positions has been proposed by \citet{CG97}.  In PIZA, a randomly chosen
tentative correspondence between initial and final positions is successively
improved by swapping randomly selected \emph{pairs} of initial particles
whenever \eqref{discrmonotone} is not satisfied. After the
cost~\eqref{discraction} ceases to decrease between iterations, an
approximation to a monotone correspondence is established, which is generally
neither unique, as already observed by \citet*{VST00b} in testing PIZA
reconstruction, nor optimal. We shall come back to this in
Sections~\ref{s:testing} and \ref{s:variational}.

\subsection{Nuts and bolts of solving the assignment problem}
\label{s:nuts}

 For a general set of $N$ unit masses, the
assignment problem with the cost function \eqref{discraction} has a
single solution which can obviously be found by examining all $N!$
permutations. However, unlike computationally hard problems, such as
the travelling salesman's, the assignment problem can be handled
in `polynomial time' -- actually in not more than $O(N^3)$
operations. All methods achieving this use a so-called dual
 formulation of the problem, based on a relaxation similar to that
applied by Kantorovich to the optimal mass transportation
(Section~\ref{s:least}; a brief introduction to duality is given in
Appendix~\ref{a:dualityproper}). In this section we explain the basics
of this technique, using a variant of a simple mechanical model
introduced in a more general setting by \citet{H95,H02}.

Consider the general assignment problem of minimizing the cost
\eqref{gendiscraction} over all permutations $j(i)$. We replace it by
a `relaxed,' linear programming problem of minimizing
\begin{equation}
  \label{relaxdiscraction}
  \tilde I = \sum_{i, j = 1}^N c_{ij} f_{ij},
\end{equation}
where auxiliary variables $f_{ij}$ satisfy
\begin{equation}
  \label{discrconstraints}
  f_{ij} \ge 0, \qquad \sum_{k = 1}^N f_{kj} = \sum_{k = 1}^N f_{ik} = 1
\end{equation}
 for all $i$, $j$, an obvious discrete analogue of \eqref{marginals}.
We show now that it is possible to build a simple mechanical device
(Fig.~\ref{f:henonmachine}) which solves this relaxed problem and that
the solution will in fact determine a minimizing permutation in the
original assignment problem (i.e., for any $i$ or~$j$ fixed, only one
$f_{ij}$ will be unit and all other zero). The device acts as an
\emph{analogue computer}: the numbers involved in the problem are
represented by physical quantities, and the equations are replaced by
physical laws.

\begin{figure}
  \iffigs
    \centerline{\psfig{file=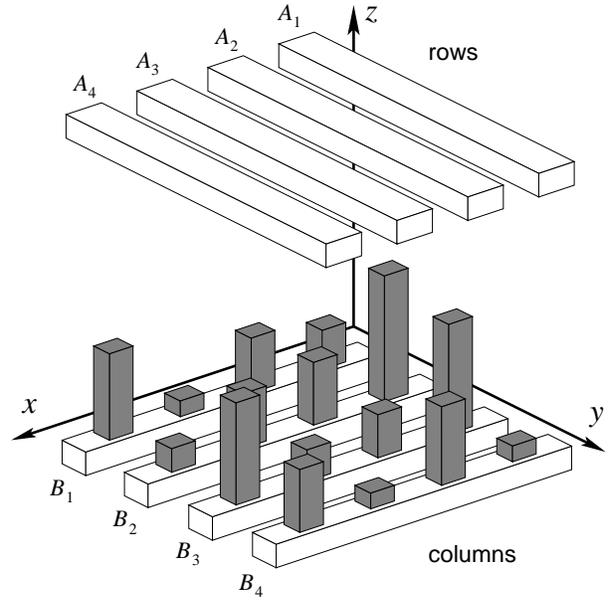,width=8cm}}
  \fi
  \vspace{2mm}
  \caption{An analogue computer solving the assignment problem for
    $N=4$.}
  \label{f:henonmachine}
\end{figure}

Define coordinate axes $x$, $y$, $z$ in space, with the $z$ axis
vertical. We take two systems of $N$ horizontal rods, parallel to the
$x$ and $y$ axes respectively, and call them \emph{columns} and
\emph{rows}, referring to columns and rows of the cost matrix. Each
rod is constrained to move in a corresponding vertical plane while
preserving the horizontal orientation in space. For a row rod $A_i$,
we denote the $z$ coordinate of its bottom face by $\alpha_i$ and for
a column rod $B_j$, we denote the $z$ coordinate of its top face
$\beta_j$. Row rods are placed above column rods, therefore $\alpha_i
\ge \beta_j$ for all $i,j$ (see Fig.~\ref{f:henonmachine}).

Upper (row) rods are assumed to have unit weight, and lower (column)
rods to have negative unit weight, or unit `buoyancy.' Therefore
both groups of rods are subject to gravitational forces pulling them
together. However, this movement is obstructed by $N^2$ small vertical
studs of negligible weight put on column rods just below row rods. A
stud placed at projected intersection of column $B_j$ and row $A_i$
has length $C-c_{ij}$ with a suitably large positive constant~$C$ and
thus constrains the quantities $\alpha_i$ and $\beta_j$ to satisfy the
stronger inequality
\begin{equation}
  \label{discrdualconstr}
  \alpha_i - \beta_j \ge C-c_{ij}.
\end{equation}
The potential energy of the system is, up to a constant,
\begin{equation}
  \label{poten}
  U = \sum_{i = 1}^N \alpha_i - \sum_{j = 1}^N \beta_j.
\end{equation}
In linear programming, the problem of minimizing \eqref{poten} under
the set of constraints given by \eqref{discrdualconstr} is called the
\emph{dual problem} to the `relaxed' one
\eqref{relaxdiscraction}--\eqref{discrconstraints} (see
Appendix~\ref{a:dualityproper}); the $\alpha$ and~$\beta$ variables
are called the \emph{dual variables}.

The analogue computer does in fact solve the dual problem. Indeed, first hold
the two groups of rods separated from each other and then release them, so
that the system starts to evolve. Rows will go down, columns will come up, and
contacts will be made with the studs. Aggregates of rows and columns will be
progressively formed and modified as new contacts are made, giving rise to a
complex evolution. Eventually the system reaches an equilibrium, in which its
potential energy \eqref{poten} is minimum and all constraints
\eqref{discrdualconstr} are satisfied \citep{H02}. Moreover, it may be shown
that the solution to the original problem
\eqref{relaxdiscraction}--\eqref{discrconstraints} is expressible in terms of
the forces exerted by the rods on each other at equilibrium and is typically a
one-to-one correspondence between the $A_i$s and the $B_j$s (for details, see
Appendix~\ref{a:henon}).

The common feature of many existing algorithms for solving the
assignment problem, which makes them more effective computationally
than the simple enumeration of all $N!$ permutations, is the use of
the intrinsically continuous, geometric formulation in terms of the
pair of linear programming problems
\eqref{relaxdiscraction}--\eqref{discrconstraints} and
\eqref{poten}--\eqref{discrdualconstr}. The mechanical device provides
a concrete model for this formulation; in fact, assignment algorithms
can be regarded as descriptions of specific procedures to make the
machine reach its equilibrium state.\footnote{This applies to
algorithms that never violate constraints \eqref{discrdualconstr}
represented by studs; all practical assignment algorithms known to us
fall within this category.} An introduction into algorithmic aspects
of solving the assignment problem, including a proof of the $O(N^3)$
theoretical bound on the number of operations, based on the Hungarian
method of \citet{K55}, may be found in \citet{PS82}.

In spite of the general $O(N^3)$ theoretical bound, various algorithms
may show very different performance when applied to a specific
optimization problem.  During the preparation of the earlier
publication \citep{FMMS02} the dual simplex method of
\citet{B86} was used, with some modifications inspired by algorithm~B
of \citet{H02}. Several other algorithms were tried subsequently,
including an adaptation of algorithm~A of the latter reference and the
algorithm of \citet{BD80}, itself based on the earlier work of
\citet{T71}. For the time being, the fastest running code by far is
based on the auction algorithm of \citet{B92a,B01}, arguably the most
effective of existing ones, which is discussed in the next section.
Needless to say, all these algorithms arrive at the same solution to
the assignment problem with given data but can differ by several
orders of magnitude in the time it takes to complete the computation.

\subsection{The auction algorithm}
\label{s:auction}

We explain here the essense of the auction algorithm in terms of our mechanical
device.%
\footnote{A movie illustrating the subsequent discussion may be found
at http://www.obs-nice.fr/etc7/movie.html (requires fast Internet 
access).} Note that the original
presentation of this algorithm \citep{B81,B92a,B01} is based on a different
perspective, that of an \emph{auction}, in which the optimal assignment
appears as an economic rather than a mechanical equilibrium; the interested
reader will benefit much from reading these papers.

Put initially the column rods at zero height and all row rods well above them,
so that no contacts are made and constraints \eqref{discrdualconstr} are
satisfied. To decrease the potential energy, let now the row rods descend
while keeping the column rods fixed. Eventually all row rods will meet studs
placed on column rods and stop. Some column rods may then come in contact
with multiple row rods. Such rods are overloaded: if they were not prevented
from moving they would descend.

Note that at this stage any column rod~$A_i$ has established a contact with
a row rod~$B_j$ for which the stud length~$C-c_{ij}$ is the maximum and the
cost $c_{ij}$ the minimum among other $B$s; for $c_{ij} = |\x_i-\q_j|^2/2$,
this means that any Eulerian position~$\x_i$ is coupled to its nearest
Lagrangian neighbour~$\q_j$. This coupling is a reasonable guess for the
optimal assignment; should it happen to be one-to-one, then the equilibrium,
and with it the optimal assignment, would be reached. It is usually not, so
there are overloaded $B$ rods and  the following procedure is applied to find
a compromise between minimization of the total cost and the requirement of
one-to-one correspondence. 

Take any overloaded rod $B_j$ and let it descend while keeping other column
rods fixed. As $B_j$ descends, row rods touching it will follow its motion
until they meet studs of other column rods and stay behind. The downward
motion of $B_j$ is stopped only when the last row rod touching $B_j$ is about
to lose its contact. We then turn to any other overloaded column rod and
repeat the procedure as often as needed.

This general step can be viewed as an \emph{auction} in which row rods
bid for the descending column rod, offering prices equal to decreases
in their potential energy as they follow its way down.  As the column
rod descends, thereby increasing its price, the auction is won by the
row rod able to offer the largest \emph{bidding increment}, i.e., to
decrease its potential energy by the largest amount while not
violating the constraints posed by studs of the rest of column rods.
 For computational purposes it suffices to compute bidding increments
 for all competing row rods from the dual $\alpha$ and $\beta$
variables and assign the descending column rod $B_j$ to the highest
bidder $A_i$, decreasing their heights $\beta_j$ and~$\alpha_i$
correspondingly.

Observe that, at each step, the total potential energy $U$ defined by
\eqref{poten} decreases by the largest amount that can be achieved by
moving the descending column rod 
without violating the constraints.\footnote{This idea of moving a rod,
or adjusting a dual variable, up to the last point compatible with all
the constraints, may be actually implemented in a number of ways,
giving rise to several possible flavours of the auction algorithm. For
example, the above procedure in its most effective implementation
requires a parallel computer so that groups of several rods can be
tracked simultaneously. On sequential computers another, less
intuitive procedure, in which upper rods are dropped once at a time,
proves more effective \citep{B92a}.}
Since \eqref{poten} is obviously
nonnegative, the descent cannot proceed indefinitely, and the process
may be expected to converge quite fast to a one-to-one pairing that
solves the assignment problem. 

However, as observed by \citet{B81,B92a,B01}, this `naive' auction
algorithm may end up in an infinite cycle if several row rods bid for
a few equally favourable column rods, having thus zero bidding
increments.  To break such cycles and also to accelerate
convergence,  a perturbation mechanism is
introduced in the algorithm.  Namely, the constraints
\eqref{discrdualconstr} are replaced by weaker ones
\begin{equation}
  \label{discrdualconstreps}
  \alpha_i - \beta_j \ge C-c_{ij} - \epsilon
\end{equation}
 for a small positive quantity $\epsilon$, and in each auction the
descending column rod is pushed down by $\epsilon$ in addition to
decreasing its height by the bidding increment.  It can be shown that
this reformulated process terminates in a finite number or rounds;
moreover, if all stud lengths are integer and $\epsilon$ is smaller
than $1/N$, then the algorithm terminates at an assignment that is
optimal in the unperturbed problem \citep{B92a}.

The third ingredient in the Bertsekas algorithm is the idea of
\emph{$\epsilon$-scaling}. When the values of dual variables are
already close to the solution of the dual problem, it usually takes
relatively few rounds of auction to converge to a solution. Thus one
can start with large $\epsilon$ to compute a rough approximation for
dual variables fast, without worrying about the quality of the
assignment, and then proceed reducing $\epsilon$ in geometric
progression until it passes the $1/N$ threshold, assuring that the
assignment thus achieved solves the initial problem.

Bertsekas' algorithm is especially fast for \emph{sparse assignment problems},
in which rods $A_i$ and $B_j$ can be matched only if the pair $(i, j)$ belongs
to a given subset $\mathcal{A}$ of the set of $N^2$ possible pairs. We call
such pairs \emph{valid} and define the \emph{filling factor} to be the
proportion of valid pairs $f = |\mathcal{A}| / N^2$. When this factor is small,
computation can be considerably faster: to find the bidding increment for a
rod~$A_i$, we need only to run over the list of rods $B_j$ such that $(i, j)$
is a valid pair.

Note also that the decentralized structure of the algorithm
 facilitates its parallelization \citep[see references in][]{B92a,B01}.

\subsection{The auction algorithm for the MAK reconstruction}
\label{s:auctionmak}

We now describe the adaptation of the auction algorithm to the MAK
reconstruction. Experiments with various programs contained in
Bertsekas' publicly available package
(http://web.mit.edu/dimitrib/www/auction.txt) showed that the most
effective for our problem is {\textsc{auction}\verb|_|\textsc{flp}}. It assumes
integer costs~$c_{ij}$, which in our case requires proper scaling of the cost
matrix. To achieve this, the unit of length is adjusted so that the size of
the reconstruction patch equals 100, and then the square of the distance
between an initial and a final position is rounded off to an integer. In our
application, row and column rods correspond to Eulerian and Lagrangian
positions, respectively. As the MAK reconstruction is planned for application
to catalogues of $10^5$ and more galaxies, we do not store the cost matrix,
which would require an $O(N^2)$ storage space, but rather compute its elements
on demand from the coordinates, which requires only $O(N)$ space.

Our problem is naturally adapted for a sparse description if galaxies
travel only a short distance compared to the dimensions of the
reconstruction patch. For instance, in the simulation discussed in
Section~\ref{s:testing}, the r.m.s. distance traveled is only about
$10\,h^{-1}$\,Mpc, or 5\% of the size of the simulation box, and the
largest distance traveled is about 15\% of this size. So we may assume
that in the optimal assignment distances between paired positions will
be limited. We define then a critical distance $d_{\rm crit}$ and
specify that a final position $\x_i$ and an initial position $\q_j$
 form a valid pair only if they are within less than $d_{\rm crit}$
 from each other. This critical distance must be adjusted carefully: if
it is too small, we risk excluding the optimal assignment; if it is
taken too large, the benefit of the sparse description is lost.

\begin{figure}
  \iffigs
    \centerline{\psfig{file=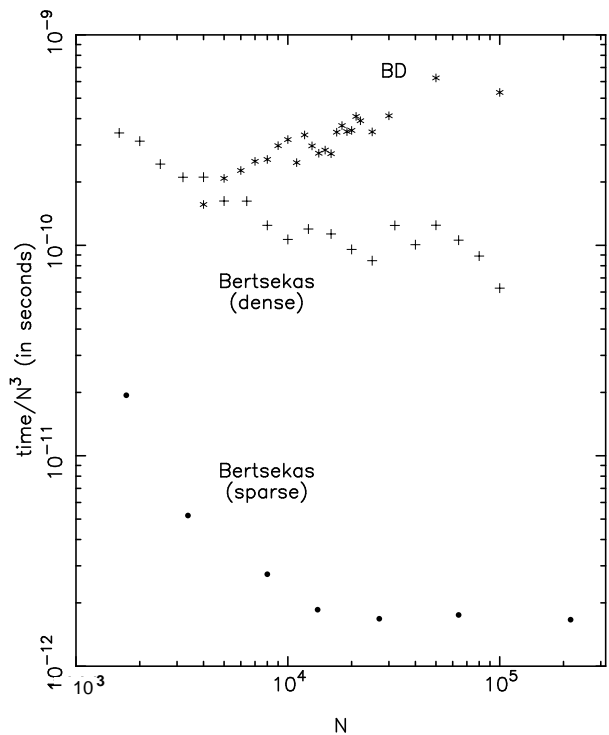,width=8cm}}
  \fi
  \vspace{2mm}
  \caption{Computing time for different algorithms as a function of
    the number $N$ of points (divided by $N^3$ for
    normalization). Asterisks, the \citet{BD80} algorithm (BD);
    crosses and points, the dense and sparse versions of the auction
    algorithm (described in the text).}
  \label{f:complexity}
\end{figure}

However, the saving in computing time achieved by sparse description has to be
paid for in storage space: to store the set $\mathcal{A}$ of valid pairs,
storage of size $|\mathcal{A}| = f N^2$ is needed, which takes us back to the
$O(N^2)$ storage requirement. We have explored two solutions to this problem.

1. Use a \emph{dense} description nevertheless, i.e.\ the one where
all pairs $(i, j)$ are valid and there is no need to store the
set~$A$. The auction program is easily adapted to this case (in fact
this simplifies the code). However, we forfeit the saving in time
provided by the sparse structure. 

2. The sparse description can be preserved if the set of valid pairs
is computed on demand rather than stored.  This is easy if initial
positions fill a uniform cubic grid, the simplest discrete
approximation to the initial quasi-uniform distribution of matter in
the reconstruction problem. Thus, for a given final position $\x_i$,
the valid pairs correspond to points of the cubic lattice that lie
inside a sphere of radius $d_{\rm crit}$ centered at $\x_i$, so their
list can be generated at run time.

 Fig.~\ref{f:complexity} gives the computing time as a function of the
number of points $N$ used in the assignment problem. Shown are the
dense and sparse versions of the auction algorithm (in the latter, the
critical distance squared was taken equal to 200) and the \citet{BD80}
algorithm, which ranked the next fastest in our experiments. The $N$
initial and final positions are chosen from the file generated by an
$N$-body simulation described in Section~\ref{s:testing}; the choice
is random except for the sparse algorithm, in which the initial
positions are required to fill a cubic lattice. Hence, the performance
of the sparse auction algorithm shown in the figure is not completely
comparable to that of the two other algorithms.

It is evident that the difference in computing time between the dense
auction and the \citeauthor{BD80} algorithms steadily increases.  In
the vicinity of $N = 10^5$, the dense auction algorithm is about 10
times faster than the other one.  For the sparse version, the decrease
in computing time is spectacular: as could be expected, the ratio of
computing times for the two versions of the auction algorithm is of
the order of~$f$. For large $N$, the $O(N^3)$ asymptotic of the computing
time is quite clear for the sparse auction algorithm. For two other
algorithms, similar asymptotic was found for larger $N$ in other
experiments (not shown).

In all three cases shown, the initial positions fill a constant volume while
$N$ is varied. This is what we call \emph{constant-volume computations}. In
the sparse case, this results in a constant filling factor, equal to the ratio
of the volume of the sphere with radius $d_\mathrm{crit}$ to the volume
occupied by the initial positions. Here this filling factor is about $f =
0.019$. Another choice, not shown in the figure, is that of
\emph{constant-density computations}, when the initial positions are taken
from a volume whose size increases with~$N$. In this case the time dependence
of algorithms for large~$N$ is of the order of~$N^{1.5}$.

We finally observe that the sparse auction algorithm applied to the
MAK reconstruction requires 5~hours of single-processor CPU time on a
667\,MHz COMPAQ/DEC Alpha machine for 216,000 points.

\section{Testing the MAK reconstruction}
\label{s:testing}

In this section we present results of our testing the MAK
reconstruction against data of cosmological $N$-body simulations. In a
typical simulation of this kind, the dark matter distribution is
approximated by $N$ particles of identical mass. Initially the
particles are put on a uniform cubic grid and given velocities that
 form a realization of the primordial velocity field whose statistics
is prescribed by a certain cosmological model. Trajectories of
particles are then computed according to the Newtonian dynamics in a
comoving frame, using periodic boundary conditions. The reconstruction
problem is therefore to recover the pairing between the initial
(Lagrangian) positions of the particles and their present (Eulerian)
positions in the $N$-body simulation, knowing only the set of computed
Eulerian positions in the physical space.

We test our reconstruction against a simulation of $128^3$ particles in a box
of $200\,h^{-1}$\,Mpc size (where $h$ is the Hubble parameter in units of
100\,km\,\,s$^{-1}$\,Mpc$^{-1}$) performed using the adaptive P$^3$M code
\textsc{HYDRA} \citep*{CTP95}.\footnote{In a flavour of $N$-body codes called
particle-mesh (PM) codes, Newtonian forces acting on particles are
interpolated from the gravitational field computed on a uniform mesh. In very
dense regions, precision is increased by adaptively refining the mesh and by
direct calculation of local particle-particle (PP) interactions; codes of this
type are correspondingly called \emph{adaptive P$^3\!$M}.} A $\Lambda$CDM
cosmological model is used with parameters $\Omega_m=0.3$,
$\Omega_\Lambda=0.7$, $h=0.65$, $\sigma_8=0.9$.\footnote{The use of a
$\Lambda$CDM model instead of the model without a cosmological constant
(Appendix~\ref{a:cosmobasics}) leads to some modifications in basic equations
but does not change formulas used for the MAK reconstruction.} The value of
these parameters within the model are determined by fitting the observed
cosmic microwave background (CMB) spectrum.\footnote{Data of the first year
Wilkinson Microwave Anisotropy Probe \citetext{\citealp{Setal03}; \citealp[see
also][]{BLOS03}} suggest a value $\sigma_8=0.84 \pm 0.04$, marginally smaller
than the one used here. This may slightly extend the range of scales
favourable for the MAK reconstruction.}  The output of the $N$-body simulation
is illustrated in Fig.~\ref{f:sky} by a projection onto the $x$-$y$ plane of a
10\% slice of the simulation box.

\begin{figure}
  \iffigs
    \centerline{\psfig{file=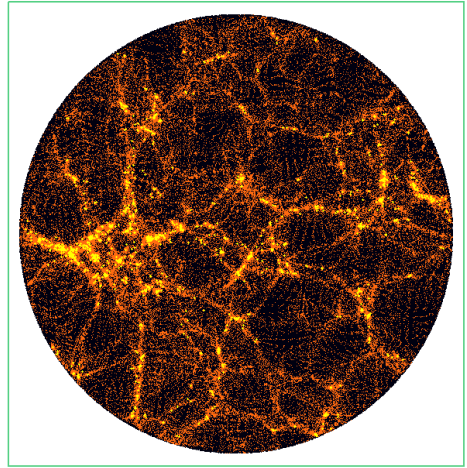,width=6.5cm}}
  \fi
  \vspace{2mm}
  \caption{$N$-body simulation output in the Eulerian space used for
    testing our reconstruction method (shown is a projection onto the
    $x$-$y$ plane of a 10\% slice of the simulation box of size
    $200h^{-1}$\,Mpc). Points are highlighted in yellow when
    reconstruction fails by more than $6.25\,h^{-1}$\,Mpc, which
    happens mostly in high-density regions.}
  \label{f:sky}
\end{figure}

Since the simulation assumes periodic boundary conditions, some
Eulerian positions situated near boundaries may have their Lagrangian
antecedents at the opposite side of the simulation box. Suppressing
the resulting spurious large displacements is crucial for successful
reconstruction. Indeed, for a typical particle displacement of 1/20
the box size, spurious box-wide leaps of 1\% of the particles will
generate a contribution to the quadratic cost \eqref{discraction} four
times larger than that of the rest. To suppress such leaps, for each
Eulerian position that has its antecedent Lagrangian position at the
other side of the simulation box, we add or subtract the box size from
coordinates of the latter (in other words, we are considering the distance
on a torus). In what follows we refer to this procedure
as the \emph{periodicity correction}.

We first present reconstructions for three samples of particles
initially situated on Lagrangian subgrids with meshes given by $\Delta
x = 6.25\,h^{-1}$\,Mpc, $\Delta x/2$ and $\Delta x/4$. To further
reduce possible effects of the unphysical periodic boundary condition,
we truncate the data by discarding 
those points whose Eulerian positions are not within the
sphere of radius $16\Delta x$ placed at the centre of the simulation
box (for the largest $\Delta x$ its diameter coincides with the box
size). The problem is then confined to finding the pairing between the
remaining Eulerian positions and the set of their
periodicity-corrected Lagrangian antecedents in the $N$-body
simulation. 

The results are shown in Figs.\ \ref{f:hist6}--\ref{f:histhuge}. The main
plots show the scatter of reconstructed vs.\ simulation Lagrangian
positions for the same Eulerian positions. 
 For these diagrams we introduce a `quasi-periodic projection'
\begin{equation}
  \tilde q \equiv 
  (q_1+\sqrt{2} q_2 + \sqrt{3}q_3)/(1 + \sqrt{2} + \sqrt{3})
\end{equation}
of the vector~$\q$, which ensures a one-to-one correspondence between
$\tilde q$-values and points on the regular Lagrangian grid. The
insets are histograms (by percentage) of distances, in reconstruction
mesh units, between the reconstructed and simulation Lagrangian
positions; the first darker bin, slightly less than one mesh in width,
corresponds to perfect reconstruction (thereby allowing a good
determination of the peculiar velocities of galaxies).

\begin{figure}
  \iffigs
    \centerline{\psfig{file=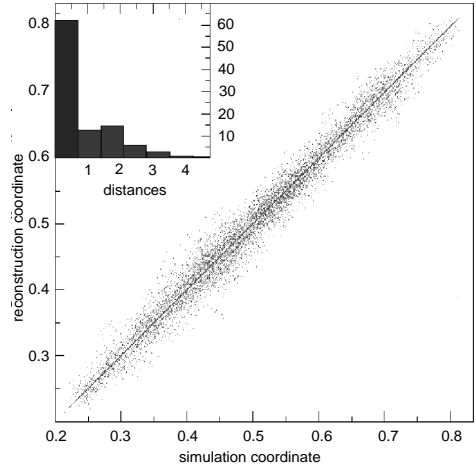,width=6.5cm}}
  \fi
    \vspace{2mm}
  \caption{Test of the MAK reconstruction for a sample of $N' =17,178$
    points initially situated on a cubic grid with mesh $\Delta x =
    6.25\,h^{-1}$\,Mpc. The scatter diagram plots true versus
    reconstructed initial positions using a quasi-periodic projection
    which ensures one-to-one correspondence with points on the cubic
    grid. The histogram inset gives the distribution (in percentages)
    of distances between true and reconstructed initial positions; the
    horizontal unit is the sample mesh. The width of the first bin is
    less than unity to ensure that only exactly reconstructed points
    fall in it. Note that more than sixty percent of the points are
    exactly reconstructed.}
  \label{f:hist6}
\end{figure}

\begin{figure}
  \iffigs
    \centerline{\psfig{file=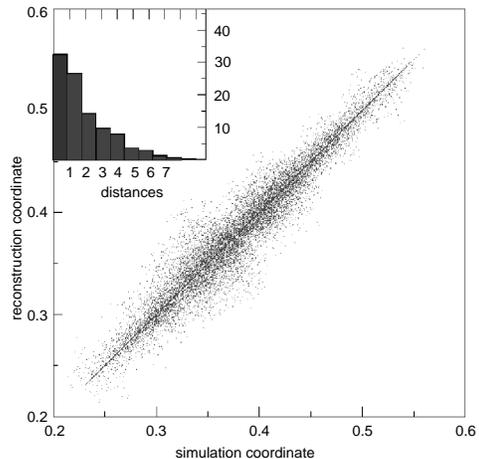,width=6.5cm}}
  \fi
  \vspace{2mm}
  \caption{Same as Fig.~\ref{f:hist6} but with $N' = 19,187$ and a
    sample mesh of $\Delta x/2 = 3.125\,h^{-1}$\,Mpc. Exact
    reconstruction is down to 35\%.}
  \label{f:hist3}
\end{figure}

\begin{figure}
  \iffigs
    \centerline{\psfig{file=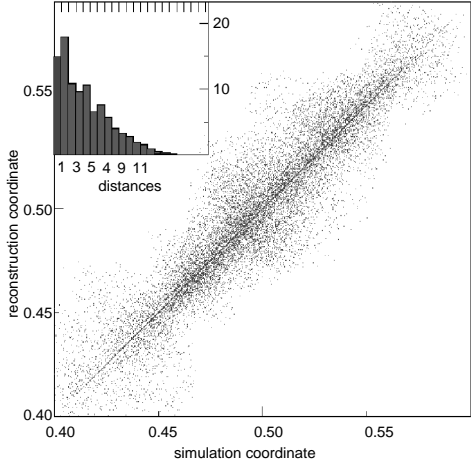,width=6.5cm}}
  \fi
  \vspace{2mm}
  \caption{Same as Fig.~\ref{f:hist6} but with $N' = 23,111$ and a
    sample mesh of $\Delta x/4 =1.56\,h^{-1}$\,Mpc. Exact
    reconstruction is down to 14\%.}
  \label{f:hist1}
\end{figure}

With the mesh size $\Delta x$, Lagrangian positions of 62\% of the
sample of 17,178 points are reconstructed perfectly and about 75\% are
placed within not more than one mesh.  With the $\Delta x/2$ grid, we
still have 35\% of exact reconstruction out of 19,187 points, but only
14\% for the $\Delta x/4$ grid with 23,111 points.

We also performed a reconstruction on a random sample of 100,000
Eulerian positions taken with their periodicity-corrected Lagrangian
antecedents out of the whole set of $128^3$ particles, without any
restrictions. This reconstruction, with the effective mesh size
(average distance between neighbouring points) of 4.35$h^{-1}$Mpc,
gives 51\% of perfect reconstruction (Fig.~\ref{f:histhuge}).

\begin{figure}
  \iffigs
    \centerline{\psfig{file=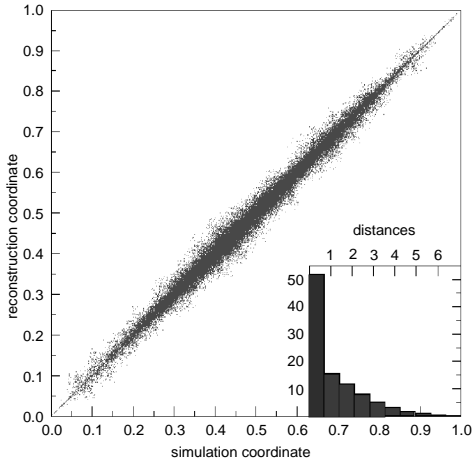,width=6.5cm}}
  \fi
  \vspace{2mm}
  \caption{Same as Fig.~\ref{f:hist6} with $N' = 10^5$ points selected
    at random, neighbouring points being typically $4.35\,h^{-1}$\,Mpc
    apart. Exact reconstruction is in excess of 50\%.}
  \label{f:histhuge}
\end{figure}

We compared these results with those of the PIZA reconstruction method
\citep[see Section~\ref{s:assignment} and][]{CG97}, which gives a
2-monotone but not necessarily optimal pairing between Lagrangian and
Eulerian positions.  We applied the PIZA method on the $\Delta x$ grid
and obtained typically 30--40\% exactly reconstructed positions, but
severe non-uniqueness: for two different seeds of the random generator
used to set up the initial tentative assignment, only about half of
the exactly reconstructed positions were the same (see figs.\ 3 and~7
of \citet{MFMS03} for an illustration). We also implemented a
modification of the PIZA method establishing 3-monotonicity
(monotonicity with respect to interchanges of 3 points instead of
pairs) and checked that it does not give a significant improvement
over the original PIZA.

In comoving coordinates, the typical displacement of a mass element is
about 1/20 the box size, that is about
$10\,h^{-1}$\,Mpc.  This is not much larger than the coarsest grid of
$6.25\,h^{-1}$\,Mpc used in testing MAK which gave 62\% of exact
reconstruction.  Nevertheless there are 18 other grid points within
$10\,h^{-1}$\,Mpc of any given grid point, so that this high
percentage cannot be trivially explained by the smallness of the
displacement.  Note that without the periodicity correction, the
percentage of exact reconstruction for the coarsest grid degraded
significantly (from 62\% to 45\%) and the resulting cost was far from
the true minimum.

 For real catalogues, reconstruction has to be performed for galaxies whose
positions are specified in the \emph{redshift space}, where they appear to be
displaced radially (along the line of sight) by an amount proportional to the
radial component of the peculiar velocity. Thus, at the present epoch, the
redshift position~$\s$ of a mass element situated at the point~$\x$ in the
physical space is given by
\begin{equation}
  \label{red}
  \s = \x + \hat\x \beta \left(\v \cdot \hat\x\right),
\end{equation}
where $\v$ is the peculiar velocity in the comoving coordinates~$\x$
and the linear growth 
factor time~$\tau$, $\hat\x$ denotes the unit normal in the direction
of~$\x$, and the parameter $\beta$ equals $0.486$ in our $\Lambda$CDM model.

 Following \citeauthor{VST00b} \citetext{\citeyear{VST00b};
\citealp[see also][]{ME99}}, we use the Zel'dovich approximation (ZA)
to render our MAK quadratric cost function in the $\s$ variable. As
 follows from \eqref{zalagmap1}, in this approximation the peculiar
velocity is given by
\begin{equation}
  \v = \frac1\tau (\x - \q).
\end{equation}
At the present time, since $\tau_0 =1$, this together with \eqref{red} 
gives
\begin{eqnarray}
  \label{redtemp1}
  (\s-\q)\cdot\hat\x&=&(1+\beta)(\x-\q)\cdot\hat\x,\\
  \label{redtemp2}
  |\s-\q|^2&=&|\x-\q|^2+\beta(\beta+2)\bigl((\x-\q)\cdot\hat\x\bigr)^2.
\end{eqnarray}
Combining now these two equations and using the fact that, by
\eqref{red}, the vectors $\x$ and~$\s$ are collinear and therefore
$\hat\x=\pm\hat\s$, we may write the quadratic cost function as
\begin{equation}
  \label{redaction}
  \frac12|\x-\q|^2 = \frac12|\s-\q|^2 - 
  \frac{\beta(\beta+2)}{2(\beta+1)^2} \bigl((\s-\q)\cdot\hat\s\bigr)^2.
\end{equation}
The redshift-space reconstruction is then in principle reduced to the
physical-space reconstruction. Note however that the redshift
transformation of Eulerian positions may fail to be one-to-one if the
peculiar component of velocity field in the proper space coordinates
exceeds the Hubble expansion component. This undermines the simple
reduction outlined above for catalogues confined to small distances.

We have performed a MAK reconstruction with the redshift-modified cost
 function \eqref{redaction}. The redshift positions were computed for
the simulation data with peculiar velocities smoothed over a sphere
with radius of 1/100 the box size ($2\,h^{-1}$\,Mpc). This
reconstruction led to 43\% of exactly reconstructed positions and 60\%
which are within not more than one $\Delta x$ mesh from their correct
positions (see Fig.~\ref{f:hist6redshift}; a scatter diagram is
omitted because it is quite similar to that in Fig.~\ref{f:hist6}). A
comparison of the redshift-space MAK reconstruction with the
physical-space MAK reconstruction shows that almost 50\% of exactly
reconstructed positions correspond to the same points.  This test
shows that the MAK method is robust with respect to systematic errors
introduced by the redshift transformation.

\begin{figure}
  \iffigs
    \centerline{\psfig{file=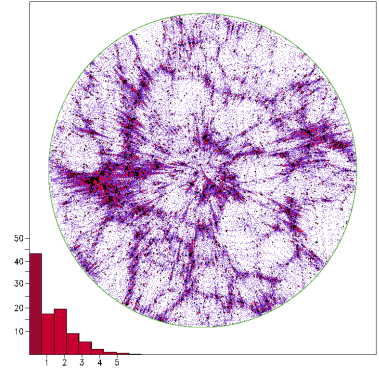,width=5cm}}
  \fi
  \vspace{2mm}
  \caption{Test of the redshift-space variant of the MAK
    reconstruction based on the same data as Fig.~\ref{f:hist6}. The
    circular redshift map (violet points) corresponds to the same
    physical-space slice as displayed in Fig.~\ref{f:sky} (the
    observer is taken at the center of the simulation
    box). Points are highlighted in red when reconstruction
    fails by more than one mesh.}
  \label{f:hist6redshift}
\end{figure}

Our results demonstrate the essentially potential character of the
Lagrangian map above $\sim 6\,h^{-1}$\,Mpc (within the $\Lambda$CDM
model) and perhaps at somewhat smaller scales.

Although it is not our intention in this paper to actually implement
the MAK reconstruction on real catalogues, a few remarks are in
order. The effect of the catalogue selection function can be handled
by standard techniques; for instance one can assign each galaxy a
`mass' inversely proportional to the catalog selection function
\citep{NB00,VST00b,BEN02}. Biasing can be taken into account in a
similar manner \citep{NB00}. Both these modifications and the natural
scatter of masses in the observational catalogues require that massive
objects be represented by clusters of multiple Eulerian points of unit
mass (with the correspondingly increased number of points on a finer
grid in the Lagrangian space), which reduces the problem to 
a variant of the usual assignment. 
We also observe that real catalogues involve truncation, that is data available
only over a finite region. As already discussed in 
Section~\ref{s:redholes},
this is not a serious problem provided a sufficiently large patch is 
available. Actually, as noted earlier in this Section, the data used in 
testing have been truncated spherically, without significantly affecting 
the quality of the reconstruction.

In the redshift-space modification, more accurate determination of
peculiar velocities can be done using second-order Lagrangian
perturbation theory.  Note also that, for the observational
catalogues, the motion of the local group itself should also be
accounted for \citep{TV99}.

\section{Reconstruction of the full self-gravitating dynamics}
\label{s:selfgravitating}

The MAK reconstruction discussed in Sections~\ref{s:makcontinuous} and
\ref{s:makdiscrete} was performed under the assumption of a potential
Lagrangian map and of the absence of multi-streaming. The tests done in
Section~\ref{s:testing} indicate that potentiality works well at scales above
$6\,h^{-1}$\,Mpc, whereas multi-streaming is mostly believed to be unimportant
above a few megaparsecs. There could thus remain a substantial range of scales
over which the quality of the reconstruction can be improved by relaxing the
potentiality assumption and using the full self-gravitating dynamics.  Here we
show that, as long as the dynamics can be described by a solution to the
Euler--Poisson equations, 
the prescription of the present density field still determines a unique
solution to the full reconstruction problem. We give only the main ideas,
technical details being left for Appendix~\ref{a:details} (a mathematically
rigorous proof may be found in \citet{L03}). In order to make the exposition
self-contained, we also give in Appendix~\ref{a:duality} an elementary
introduction to convexity and duality which are used for the derivation (and
also elsewhere in this paper).

We shall start from an Eulerian variational formulation of the
Euler--Poisson equations in an Einstein--de~Sitter universe, which is
an adaptation of a variational principle given by \citet{GMMY93}. We
minimize the action
\begin{equation}
  \label{action}
  I = \frac12 \int_0^{\tau_0}\d\tau\int  \D\x\,
  \tau^{3/2}\left(\rho|\v|^2 + \frac32|\gradx\phig|^2\right),
\end{equation}
under the following four constraints: the Poisson equation
\eqref{repcomovpoisson}, the mass conservation equation
\eqref{repcomovcontinuity} and the boundary conditions that the density field
be unity at $\tau =0$ and prescribed at the present time $\tau =\tau_0$.  The
constraints can be handled by the standard method of Lagrange multipliers
(here functions of space and time), which allows to vary independently the
 fields $\rho$, $\phig$ and $\v$.  The vanishing of the variation
in~$\v$ gives $\v = \tau^{-3/2}\gradx\theta$,  where $\theta(\x,\tau)$ is
the Lagrange multiplier  for the mass conservation constraint. Hence,
the velocity is curl-free. The vanishing of the variation in~$\rho$ gives
then 
\begin{equation}
  \dtau\theta + \frac1{2\tau^{3/2}}|\gradx\theta|^2 
  + \frac3{2\tau}\psi = 0.
\end{equation}
By taking the gradient, this equation goes over into the momentum
equation \eqref{repcomoveuler}, repeated here for convenience:
\begin{equation}
  \label{reprepcomoveuler}
  \dtau\v+(\v\cdot\gradx)\v 
  = -\frac{3}{2\tau}(\v+\gradx\phig).
\end{equation}

It is noteworthy that, if in the action we replace $3/2$ both in the
exponent of $\tau$ and in the gravitational energy term by
$3\alpha/2$, we obtain \eqref{reprepcomoveuler} but also with a
$3\alpha/(2\tau)$ factor in the right-hand side. The Zel'dovich
approximation and the associated MAK reconstruction amount clearly to setting 
$\alpha =0$, so as to recover the `free-streaming action'
\begin{equation}
  \label{zaaction}
  I = \frac12 \int_0^{\tau_0}\d\tau\int  \D\x\;
  \rho|\v|^2,
\end{equation}
whose minimization is easily shown to be equivalent to that of the 
quadratic cost function \eqref{quadraticcost}.

Assuming the action \eqref{action} to be finite, existence of a minimum is
mostly a consequence of the action being manifestedly non-negative. Here it is
interesting to observe that the Lagrangian, which is the \emph{difference}
between the kinetic energy and the potential energy, is positive whereas the
Hamiltonian which is their \emph{sum} does not have a definite sign. As a
consequence, our two-point boundary problem is, as we shall see, well posed
but the initial-value problem for the Euler--Poisson system is not well posed
since formation of caustics after a finite time cannot be ruled
out.\footnote{If we had considered electrostatic repulsive interactions the
conclusions would be reversed.}

Does the variational formulation imply uniqueness of the solution?
This would be the case if the action were a strictly convex functional
(see Appendix~\ref{a:convexity}), which is guaranteed to have one and
only one minimum. The action as written in \eqref{action} is not
convex in the $\rho$ and $\v$ variables, but can be rendered so by
introducing the mass flux $\J = \rho\,\v$; the kinetic energy term
becomes then $|\J|^2/(2\rho)$, which is convex in the $\J$ and~$\rho$
variables.

Strict convexity is particularly cumbersome to establish, but there is
an alternative way, known as duality: by a Legendre-like
transformation the variational problem is carried into a dual problem
written in terms of dual variables; the minimum value for the original
problem is the maximum for the dual problem. It turns out that the
difference of these equal values can be rewritten as a sum of
non-negative terms, each of which must thus vanish. This is then used
to prove (i) that the difference between any two solutions to the
variational problem vanishes and (ii) that any curl-free solution to
the Euler--Poisson equations with the prescribed boundary conditions
 for the density also minimizes the action. All this together
establishes uniqueness. For details see Appendix~\ref{a:details}.

Several of the issues raised in connection with the MAK reconstruction
appear in almost the same form for the Euler--Poisson
reconstruction. First, we are faced again with the problem that, when
reconstructing from a finite patch of the present universe, we need
either to know the shape of the initial domain or to make some
hypothesis as to the present distribution of matter outside this
patch.  Second, just as for the MAK reconstruction, the
proof of uniqueness still holds when the present density $\rho_0(\x)$
has a singular part, that is, when some matter is concentrated. Again,
we shall have full information on the initial shape of collapsed
regions but not on the initial fluctuations inside them. The
particular solution obtained from the variational formulation is the
only solution which stays smooth for all times prior to $\tau_0$.

We also note that, at this moment and probably for quite some time, 3D
catalogues sufficiently dense to allow reconstruction will be limited
to fairly small redshifts.  Eventually, it will however become of
interest to perform reconstruction `along our past light-cone' with
data not all at $\tau_0$. The variational approach can in principle be
adapted to handle such reconstruction.

In previous sections we have seen how to implement reconstruction using MAK,
which is equivalent to using the simplified action
\eqref{zaaction}. Implementation using the full Euler--Poisson action
\eqref{action} is mostly beyond the scope of this paper, but we shall indicate
some possible directions. In principle it should be possible to adapt to the
Euler--Poisson reconstruction the method of the augmented Lagrangian which has
been applied to the two-dimensional Monge--Amp\`ere equation \citep{BB00}.
An alternative strategy, which allows reduction to MAK-type problems, uses the
idea of `kicked burgulence' \citep*{BFK00} in which, in order to solve
the one or multi-dimensional Burgers equation 
\begin{equation}
  \dtau\v+(\v\cdot\gradx)\v = \bmath{f}(\x,\tau), \quad \v= -\gradx\phiv,
  \label{burgersf}
\end{equation}
one approximates the force by a sum of delta-functions in time:
\begin{equation}
  \bmath{f}(\x,\tau) \approx \sum_i \delta(\tau -\tau_i)\bmath{g}_i(\x).
  \label{deltakick}
\end{equation}
In the present case, the $\bmath{g}_i(\x)$ are proportional to the
right-hand side of \eqref{reprepcomoveuler} evaluated at the kicking
times $\tau_i$.  The action becomes then a sum of free-streaming
Zel'dovich-type actions plus discrete gravitational contributions
stemming from the kicking times. Between kicks one can use our MAK
solution.  At kicking times the velocity undergoes a discontinuous
change which is related to the gravitational potential (and thus to
the density) at those times. The densities at kicking times can be
determined by an iterative procedure.  The kicking strategy also
allows to do redshift-space reconstruction by applying the
redshift-space modified cost (Section~\ref{s:testing}) at the last
kick.

\section{Comparison with other reconstruction methods}
\label{s:other}

Reconstruction started with Peebles' (1989) work, in which he compared
reconstructed and measured peculiar velocities for a small number of
Local Group galaxies, situated within a few Mpc.  The focus of
reconstruction work has now moved to tackling the rapidly growing
large 3D surveys \citep[see, e.g.][]{FS00}. It is not our intention
here to review all the work on reconstruction;\footnote{For a
comparison of six different techniques, see \citet{NC99}.} rather we
shall discuss how some of the previously used methods can be
reinterpreted in the light of the optimization approach to
reconstruction.  For convenience we shall divide methods into
perturbative (Section~\ref{s:perturbative}), probabilistic
(Section~\ref{s:probabilistic}), and variational
(Section~\ref{s:variational}). Methods such as POTENT \citep{DBF90},
whose purpose is to obtain the full peculiar velocity field from its
radial components using the (Eulerian) curl-free property, are not
directly within our scope. Note that in its original Lagrangian form
\citep{BD89,DBF90} POTENT was assuming a curl-free velocity in
Lagrangian coordinates, an assumption closely related to the potential
assumption made for MAK, as already pointed out in
Section~\ref{s:potential}. Even closer is the relation
between MAK and the PIZA method of \citet{CG97}, discussed in
Section~\ref{s:variational}, which is also based on minimization of
quadratic action.

\subsection{Perturbative methods}
\label{s:perturbative}

\citet{ND92} have proposed using the Zel'dovich approximation
backwards in time to obtain the initial velocity fluctuations and thus
(by slaving) the density fluctuations.  Schematically, their procedure
involves two steps: (i) obtaining the present potential velocity field
and (ii) integrating the Zel'dovich--Bernouilli equation back in
time. Using the equality (in our notation) of the velocity and
gravitational potentials, they point out that the velocity potential
can be computed from the present density fluctuation field by solving
the Poisson equation. This is a perturbative approximation to
reconstruction in so far as it replaces the Monge--Amp\`ere equation
\eqref{MA} by a linearized form. Indeed, when using the Zel'dovich
approximation we have $\q=\x-\tau\v = \x +\tau\gradx \phiv(\x)$.
We know that $\q = \gradx\Theta(\x)$ with $\Theta$ satisfying the
Monge--Amp\`ere equation. The latter can thus be rewritten as
\begin{equation}
  \det \left(\delta_{ij} + \tau \nabla_{x_i}\nabla_{x_j}\phiv(\x)\right) =
  \rho(\x), 
  \label{MAphiv}
\end{equation}
where $\delta_{ij}$ denotes the identity matrix.  If we now use the relation
$\det (\delta_{ij}+\epsilon A_{ij})= 1+\epsilon \sum_i A_{ii} + O(\epsilon ^2)$
and truncate the expansion at order $\epsilon$, we obtain the Poisson equation
\begin{equation}
  \tau \lapx \phiv(\x) = \rho(\x) -1 = \delta(\x).
  \label{MApoisson}
\end{equation}
Of course, in one dimension no approximation is needed.  From a
physical point of view, equating the velocity and gravitational
potentials at the present epoch amounts to using the Zel'dovich
approximation in reverse and is actually inconsistent with the forward
Zel'dovich approximation: the slaving which makes the 
two potentials
equal initially does not hold in this approximation at later epochs.
Replacing the Monge--Amp\`ere equation by the Poisson equation is not
consistent with a uniform initial distribution of matter and will in
general lead to spurious multi-streaming in the initial
distribution. Of course, if the present-epoch velocity field happens
to be known one can try applying the Zel'dovich approximation in
reverse. Nusser and Dekel observe that calculating the inverse
Lagrangian map by $\q=\x-\tau\v$ does not work well (spurious
multi-streaming appears) and instead integrate back in time the
Zel'dovich--Bernouilli equation\footnote{In the non-cosmological
literature this equation is usually called Hamilton--Jacobi in the
context of analytical mechanics \citep{LL60} and Kardar--Parisi--Zhang
\citeyearpar{KPZ86} in condensed matter physics.}
\begin{equation}
  \partial_t \phiv={1\over2}\left(\gradx\phiv\right)^2,
  \label{ZB}
\end{equation}
which is obviously equivalent to the Burgers equation~\eqref{3dburgers} with
the viscosity $\nu=0$.  One way of performing this reverse integration, which
guarantees the absence of multi-streaming, is to use the Legendre
transformation~\eqref{legendre2} to calculate $\Phi(\q)$ from
$\Theta(\x)=|\x|^2/2 -\tau\phiv(\x)$ and then obtain the reconstructed initial
velocity field as
\begin{equation}
  \v_{\rm in}(\q) = \v_0\left(\gradq\Phi(\q)\right).
  \label{legendrereconstuct}
\end{equation}
This procedure can however lead to spurious shocks in the reconstructed
initial conditions, due to inaccuracies in the present-epoch velocity data,
unless the data are suitably smoothed. Finally, the improved reconstruction
method of \citet{G93} can be viewed as an approximation to the Monge--Amp\`ere
equation beyond the Poisson equation which captures part of the nonlinearity.

\subsection{Probabilistic methods}
\label{s:probabilistic}
 
\citet{W92} presents an original approach to reconstruction, which
turns out to have hidden connections to optimal mass
transportation. The key observations in his `Gaussianization'
technique are the following: (i) the initial density fluctuations are
assumed to be Gaussian, (ii) the rank order of density values is
hardly changed between initial and present states, (iii) the bulk
displacement of large-scale features during dynamical evolution can be
neglected. Assumption (i) is part of the standard cosmological
paradigm. Assumption (iii) can of course be tested in $N$-body
simulations. As we have seen in Section~\ref{s:testing}, a
displacement of 10\,$h^{-1}$Mpc is typical and can indeed be
considered small compared to the size of the simulation boxes
($64\,h^{-1}$Mpc in Weinberg's simulations and $200\,h^{-1}$Mpc in
ours). Assumption (ii) means that the correspondence between initial
and present values of the density $\rho$ (or of the contrast $\delta
=\rho -1$) is monotone. This map, which can be determined from the
empirical present data, can then be applied to all the data to produce
a reconstructed initial density field. Finally, by running an $N$-body
simulation initialized on the reconstructed field one can test the
validity of the procedure, which turns out to be quite good and can be
improved further by hybrid methods \citep{NW98,KDGW96} combining
Gaussianization with the perturbative approaches of \citet{ND92} or
\citet{G93}.

This technique is actually  connected with mass transportation:
starting with the work of Fr{\'e}chet \citetext{\citeyear{F57a,F57b};
\citealp[see also][]{R84}}, probabilists have been asking the
 following question: given two random variables $m_1$ and $m_2$ with
two laws, say PDFs $p_1$ and $p_2$, can one find a joint distribution
of $(m_1,m_2)$ with PDF $p_{12}(m_1,m_2)$ having the following
properties: (i) $p_1$ and $p_2$ are the marginals, i.e. when $p_{12}$
is integrated over $m_2$ (respectively, $m_1$) one recovers $p_1$
(respectively, $p_2$), (ii) the correlation $\langle m_1 m_2\rangle$
is maximum?  Since $\langle m_1^2\rangle$ and $\langle m_2^2\rangle$
are obviously prescribed by the constraint that we know $p_1$ and
$p_2$, maximizing the correlation is the same as minimizing the
quadratic distance $\langle (m_1-m_2)^2\rangle$. This is precisely an
instance of the mass transportation problem with quadratic cost, as we
defined it in Section~\ref{s:least}. As we know, the optimal solution
is obtained by a map from the space of $m_1$ values to that of $m_2$
values which is the gradient of a convex function. If $m_1$ and $m_2$
are scalar variables, the map is just monotone, as in the
Gaussianization method (in the discrete setting this was already
observed in Section~\ref{s:assignment}). Hence Weinberg's method may
be viewed as requiring maximum correlation (or minimum quadratic
distance in the above sense) between initial and present distributions
of density fluctuations.

In principle the Gaussianization method can be extended to multipoint
distributions, leading to a difficult multidimensional mass
transportation problem which can be discretized into an assignment
problem just as in Section~\ref{s:assignment}. The contact of the
maximum correlation assumption to the true dynamics is probably too
 flimsy to justify using such heavy machinery.

\subsection{Variational methods}
\label{s:variational}

All variational approaches to reconstruction, starting with 
that of \citet{P89}, have common features: one uses a suitable
Lagrangian and poses a two-point variational problem with boundary
conditions prescribed at the present epoch by the observed density
 field, and at early times by requiring a quasi-uniform distribution of
matter (more precisely, as we have seen in Section~\ref{s:early}, by
requiring that the solutions not be singular as $\tau\to 0$).

The Path Interchange Zel'dovich Approximation (PIZA) method of
\citet{CG97} and our MAK reconstruction techniques use a
free-streaming Lagrangian in linear growth rate time. As we have seen
in Section~\ref{s:potential}, this amounts to assuming adhesion
dynamics. Once discretized for numerical purposes, the variational
problem becomes an instance of the assignment problem.  \citet{CG97}
have proposed a restricted procedure for solving it, which
does not account for the Lagrangian potentiality 
and yields non-unique approximate solutions. As we have
seen in Sections \ref{s:makdiscrete} and~\ref{s:testing}, the exact
and unique solution can be found with reasonable CPU resources.

Turning now to the Peebles least action method, let us first describe
it schematically, using our notation. In its original formulation it
is applied to a discrete set of galaxies (assumed of course to trace
mass) in an Einstein--de~Sitter universe. The action, in our notation,
can be written as
\begin{eqnarray}
  \nonumber
   I &=&\displaystyle \int_0^{\tau_0} \d\tau\, \frac3{2\tau^{1/2}} 
  \left(\sum_i \frac{m_i\tau^2}{3}\left|\frac{d\x_i}{d\tau}\right|^2\right.\\
  \label{peeblesaction}
  &&\displaystyle\left.
   + \frac{3G}{2}\sum_{i \neq j}\frac{m_i m_j}{|\x_i - \x_j|}
   + \upi G\bar\varrho_0 \sum_i m_i |\x_i|^2\right),
\end{eqnarray}
where $m_i$ is the mass and $\x_i$ the comoving coordinate of $i$th
galaxy \citep[see also][]{NB00}.  This is supplemented by the boundary
condition that the present positions of the galaxies are known and
that the early-time velocities satisfy\footnote{This condition, which
is written $a^2d\x_i/dt \to 0$ in Peebles' notation, ensures the
vanishing of the corresponding boundary term after an integration by
parts in the time variable.}
\begin{equation}
  \tau^{3/2}\frac{d\x_i}{d\tau} \to 0 \qquad {\rm for}\,\,\,\, \tau\to 0.
  \label{shorttime}
\end{equation} 
This particle approach was extended by \citet{GMMY93} to a continuous
distribution in Eulerian coordinates and leads then to the action
analogous to \eqref{action} which we have used in
Section~\ref{s:selfgravitating}. The procedure also involves a
`Galerkin truncation' of the particle trajectories to finite sums of
trial functions of the form
\begin{eqnarray}
  x_i^\mu(\tau) &=& x_i^\mu(\tau_0) + \sum_{n=0}^{N-1} C^\mu_{i,n} f_n(\tau),
  \label{peeblesansatz}\\
  f_n(\tau) &=& \tau^n (\tau_0-\tau), \quad n=0,1,\ldots,N-1.
  \label{powers}
\end{eqnarray}
The reconstructed peculiar velocities for the Local Group were used by
Peebles to calibrate the Hubble and density parameters, which turned
out to differ from the previously assumed values. However the peculiar
velocity of one dwarf galaxy, N6822, failed to match the observed
value (see Fig.~\ref{f:localgroup}).
\begin{figure}
  \iffigs
    \centerline{\psfig{file=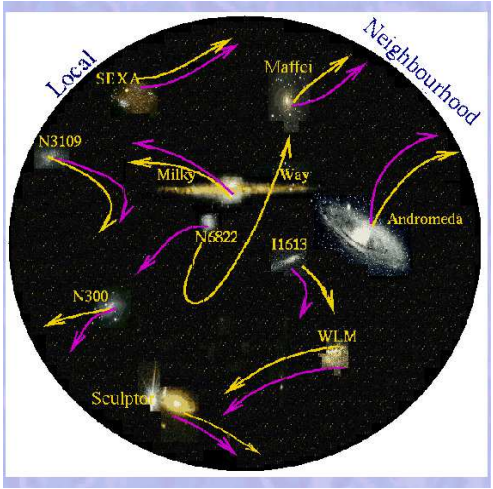,width=6.5cm}}
  \fi
  \vspace{2mm}
  \caption{A schematic demonstration of Peebles' reconstruction of the
    trajectories of the members of the local neighbourhood using a
    variational approach based on the minimization of Euler--Lagrange
    action.  The arrows go back in time, starting from the present and
    pointing towards the initial positions of the sources.  In most
    cases there is more than one allowed trajectory due to orbit
    crossing (closely related to the multi-streaming of the underlying
    dark matter fluid). The pink (darker) orbits correspond to taking
    the minimum of the action whereas the yellow (brighter) orbits
    were obtained by taking the saddle-point solution.  Of particular
    interest is the orbit of N6822 which in the former solution is on
    its first approach towards us and in the second solution is in its
    passing orbit.  A better agreement between the evaluated and
    observed velocities was shown to correspond to the saddle-point
    solution.}
  \label{f:localgroup}
\end{figure}
This led \citet{P90} to partially relax the assumption of
\emph{minimum} action, allowing also for saddle points in the
action. Somewhat better agreement with observations is then obtained,
but at the expense of lack of uniqueness.

In the context of the present approach, various remarks can be made.
The boundary condition \eqref{shorttime} is trivially satisfied if the
velocities $d\x/d\tau$ remain bounded. Actually, we have seen in
Section~\ref{s:early} that, as a consequence of slaving, the velocity
has a regular expansion in powers of $\tau$, which implies its
boundedness as $\tau \to 0$. The important point is that the function
$f_n(\tau)$ appearing in \eqref{peeblesansatz} should be expandable in
powers of $\tau$, as is the case with the ansatz \eqref{powers}.

In Section~\ref{s:selfgravitating} we have established uniqueness of
the reconstruction with a prescribed present density and under the
assumption of absence of multi-streaming (but we allow for 
mass concentrations). This restriction is meaningful only in the
continuous case: in the discrete case, unless the particles are rather
closely packed, the concept of multi-streaming is not clear but there
have been attempts to relate uniqueness to absence of `orbit crossing'
\citep[see, e.g.,][]{GMMY93,W00}. Of course, at the level of the
underlying dark matter, multi-streaming is certainly not ruled out at
sufficiently small scales; at such scales unique reconstruction is not
possible.

In the truly discrete case, e.g.\ when considering a dwarf galaxy, there is no
reason to prefer the true minimum action solution over any other stationary
action solution.

\section{Conclusion}
\label{s:conclusion}

The main theoretical result of this paper is that reconstruction of
the past dynamical history of the Universe, knowing only the present
spatial distribution of mass, is a well-posed problem with a unique
solution. More precisely, reconstruction is uniquely defined down to
those scales, a few megaparsecs, where multi-streaming becomes
important. The presence of concentrated mass in the form of clusters,
 filaments, etc is not an obstacle to a unique displacement
reconstruction; the mass within each such structure originates from a
collapsed region of known shape but with unknown initial density and
velocity fluctuations inside.  There are of course practical
limitations to reconstruction stemming from the knowledge of the
present mass distribution over only a limited patch of the Universe;
these were discussed in Section~\ref{s:redholes}.

In this paper we have also presented in detail and tested a
reconstruction method called MAK which reduces reconstruction to an
assignment problem with quadratic cost, for which 
effective algorithms are available. MAK, which is exact for dynamics
governed by the adhesion model, works very well above $6\,h^{-1}$\,Mpc
and can in principle be adapted to full Euler--Poisson reconstruction.

We note that a very common method for testing ideas about the early Universe
is to take some model of early density fluctuations and then run an $N$-body
simulations with assumed cosmological parameters until the present
epoch. Confrontation with the observed \emph{statistical properties} of the
present Universe helps then in selecting plausible models and in narrowing the
choice of cosmological parameters. This \emph{forward method} is conceptually
very different from reconstruction; the latter not only works backward but,
more importantly, it is a \emph{deterministic} method which gives us a
detailed map of the early Universe and how it relates to the present one.
Reconstruction thus allows us to obtain the peculiar velocities of galaxies
and is probably the only method which can hope to do this for a large number
of galaxies. In those instances were we have partial information on peculiar
velocities (from independent distance measurements), e.g.\ for the NearBy
Galaxies (NBG) catalogue of \citet{T88}, such information can be used to
calibrate cosmological parameters or to provide additional constraints, which
are in principle redundant but can improve the quality. 

The detailed reconstruction of early density fluctuations, which will become
possible using large 3D surveys such as 2dF and SDSS \citep[see,
e.g.,][]{FS00}, will allow us to test such assumptions as the Gaussianity of
density fluctuations at decoupling. Note however that such reconstruction
gives us full access only to the complement of collapsed regions; any
statistical information thus obtained will be biased, roughly by
overemphasizing underdense regions.

 Finally we have no reason to hide the pleasure we experience in seeing
this heavenly problem bring together and indeed depend crucially on so
many different areas of mathematics and physics, from fluid dynamics
to Monge--Amp\`ere equations, mass transportation, convex geometry and
combinatorial optimization. Probably this is the first time that one
tackles the three-dimensional Monge--Amp\`ere equation numerically for
practical purposes. As usual, we can expect that the techniques, here
applied to cosmic reconstruction, will find many applications, for
example to the optimal matching of two holographic or tomographic
images or to the correction of images in multi-dimensional colour
space.

\section*{Acknowledgments}

Special thanks are due to E.~Branchini (observational and conceptual
aspects) and to D.~Bertsekas (algorithmic aspects).  We also thank
J.~S.~Bagla, J.~Bec, E.~Bertschinger, T.~Buchert, A.~Dom{\'\i}nguez,
H.~Frisch, J.~Gaite, C.~l'Hostis, L.~Moscardini, A.~Noullez, M.~Rees,
V.~Sahni, S.~Shandarin, A.~Shnirelman, E.~Slezak, E.~Spiegel,
A.~Starobinsky, P.~Thomas and B.~Villone for comments and useful
information.

This work was supported by the European Union under contract
HPRN-CT-2000-00162, by the BQR program of Observatoire de la C\^ote d'Azur, by
the TMR program of the European Union (UF), by MIUR (SM), by the French
Ministry of Education, the McDonnel Foundation, and the Russian Foundation
for Basic Research under grant
RFBR 02-01-1062 (AS). RM
was supported by 
a Marie Curie Fellowship
HPMF-CT-2002-01532.

\newpage

\appendix 

\renewcommand{\theequation}{\thesection.\arabic{equation}}
\setcounter{equation}{0}

\section{Equations of motion in an expanding universe}
\label{a:cosmobasics}

On distances covered by present and forthcoming redshift galaxy
catalogues, the Newtonian description constitutes a realistic
approximation to the dynamics of self-gravitating cold dark matter
 filling the Universe \citep{P80,CL95}. This description gives, in
proper space coordinates denoted here by~$\r$ and cosmic time~$t$, the
 familiar Euler--Poisson system for the density~$\varrho(\r,t)$,
velocity~$\U(\r,t)$ and the gravitational potential~$\phi(\r,t)$:
\begin{eqnarray}
  \label{propereuler}
  \partial_t\U + (\U \cdot \nabla_\r)\U &=& -\nabla_r\phi_{\mathrm{g}}, \\
  \label{propercontinuity}
  \partial_t\varrho + \nabla_\r\cdot(\varrho\U) &=& 0, \\
  \label{properpoisson}
  \nabla^2_\r \phi_{\mathrm{g}} &=& 4\upi G \varrho,
\end{eqnarray}
where $G$ is the gravitation constant.

In a homogeneous isotropic universe, the density and velocity fields
take the form
\begin{equation}
  \label{homogeneousansatz}
  \varrho(\r, t) = \bar\varrho(t),\quad
  \U(\r, t) = H(t)\r = \frac{{\dot a}(t)}{a(t)}\r.
\end{equation}
Here the coefficient $H(t)$ is the \emph{Hubble parameter}, and $a(t)$
is the \emph{expansion scale factor} defined so that integration of
the velocity field $\dot\r = \U(\r, t) = H(t)\r$ yields $\r = a(t)\x$,
where $\x$ is called the \emph{comoving coordinate}.

The \emph{background density}~$\bar\varrho(t)$ gives rise to the
background gravitational potential~$\bar\phi_{\mathrm{g}}$, which by
\eqref{propereuler} and \eqref{homogeneousansatz} satisfies
\begin{equation}
  \label{tempgrav}
  -\nabla_\r \bar\phi_{\mathrm{g}} = \frac{\ddot a}{a}\r.
\end{equation}
 For the background density, mass conservation \eqref{propercontinuity}
gives then 
\begin{equation}
  \label{massconsintegr}
  \bar\varrho a^3 = \bar\varrho_0,
\end{equation}
where $\bar\varrho_0 = \bar\varrho(t_0)$ with $t_0$ the present epoch
and $a(t_0)$ is normalized to
unity. Eqs.~\eqref{tempgrav}, \eqref{massconsintegr}, and
\eqref{properpoisson} imply the \emph{Friedmann equation} for $a(t)$:
\begin{equation}
  \label{friedmann}
  \ddot a = -\frac43 \upi G \bar\varrho_0\frac{1}{a^2}
\end{equation}
with conditions posed at $t = t_0$:
\begin{equation}
  \label{friedmannini}
  a(t_0) = 1, \quad \dot a(t_0) = H_0 > 0,
\end{equation}
where $H_0$ is the present value of the Hubble parameter, positive for
an expanding universe.

 For simplicity we restrict ourselves to the case of the \emph{critical
density}, corresponding to the flat, matter-dominated Einstein-de
Sitter universe (without a cosmological constant):
\begin{equation}
  \label{rhobar}
  \bar\varrho_0 = \frac{3H_0^2}{8\upi G}
\end{equation}
and adjust the origin of the time axis such that the solution takes
the form of a power law
\begin{equation}
  \label{a}
  a(t) = \left(\frac{t}{t_0}\right)^{2/3}
\end{equation}
with $H_0 = 2/(3t_0)$ and $\bar\varrho_0 = 1/(6\upi G t_0^2)$.

The observed Hubble expansion of the Universe suggests that the
density, velocity and 
gravitational fields may be decomposed into a sum of
terms describing the uniform expansion and fluctuations against the
background:
\begin{equation}
  \varrho = \bar\varrho(t)\, \rho,\quad
  \U = \frac{\dot a(t)}{a(t)}\r + a(t)\u,\quad
  \phi_{\mathrm{g}} = \bar\phi_{\mathrm{g}} + \widetilde\phig.
\end{equation}
The term~$a(t)\u$ is called the \emph{peculiar velocity}.  In
cosmology, one also often employs the \emph{density contrast} defined
as $\delta = \rho - 1$, which gives the fluctuation against the
normalized background density.  Taking $\rho$, $\u$,
and~$\widetilde\phig$ as functions of the comoving coordinate $\x =
\r/a(t)$ and using \eqref{tempgrav}, \eqref{massconsintegr} and
\eqref{friedmann}, we rewrite the Euler--Poisson system in the form
\begin{eqnarray}
  \label{costimeeuler}
  \partial_t\u + (\u \cdot \gradx)\u 
  &=& -2\frac{\dot a}{a}\u - \frac{1}{a}\gradx\widetilde\phig,\\
  \label{costimecontinuity}
  \partial_t\rho + \gradx\cdot(\rho\u) &=& 0,\\
  \label{costimepoisson}
  \lapx\widetilde\phig &=& \frac{4\upi G\bar\varrho_0}{a}(\rho - 1).
\end{eqnarray}
Note the \emph{Hubble drag} term $-2(\dot a/a)\u$ in the right-hand side of
\eqref{costimeeuler} representing the relative slowdown of peculiar
velocities due to the uniform expansion.

 Formally linearizing \eqref{costimeeuler}--\eqref{costimepoisson}
around the trivial zero solution, one obtains the following ODE for
the \emph{linear growth factor}~$\tau(t)$ of density fluctuations:
\begin{equation}
  \label{lingrowth}
  \frac{d}{dt}(a^2\dot\tau) = 4\upi G\bar\varrho_0\frac{\tau}{a}.
\end{equation}
The only solution of this equation that stays bounded (indeed, vanishes) at
small times is usually referred to as the \emph{growing mode}. As we shall
shortly see, it is convenient to choose the amplitude factor~$\tau$ of the
growing mode to be a new `time variable,' which in an Einstein--de~Sitter
universe is proportional to~$t^{2/3}$. It is normalized such that $\tau_0 =
\tau(t_0) = 1$. Rescaling the peculiar velocity and the gravitational
potential according to
\begin{equation}
  \u = \dot\tau\v,\quad 
  \widetilde\phig = \frac{4\upi G\bar\varrho_0\tau}{a}\phig
\end{equation}
and using the fact that in an Einstein--de~Sitter universe
$d\ln(a^2\dot\tau)/d\tau = 3/(2\tau)$, we arrive at the following form
of the \emph{Euler--Poisson system}, which we use throughout this
paper:
\begin{eqnarray}
  \label{comoveuler}
  \dtau\v+(\v\cdot\gradx)\v 
  &=& -\frac{3}{2\tau}(\v+\gradx\phig),\\
  \label{comovcontinuity}
  \dtau\rho+\gradx\cdot(\rho\v) &=& 0,\\
  \label{comovpoisson}
  \lapx\phig &=& \frac{\rho - 1}{\tau}.
\end{eqnarray}

Suppose initially, i.e.\ at $\tau = 0$, a mass element is located at a
point with the comoving coordinate~$\q$. Transported by the peculiar
velocity field in the comoving coordinates, this element describes a
trajectory $\x(\q, \tau)$.  Using the \emph{Lagrangian
coordinate}~$\q$ to parametrize the whole continuum of mass elements,
we recast \eqref{comoveuler} and~\eqref{comovpoisson} in the form
\begin{eqnarray}
  \DDtau\x&=&-\frac{3}{2\tau}\left(\Dtau\x+\gradx\phig\right),
  \label{lagnewton}\\
  \lapx\phig&=&\frac{1}{\tau}\left[\left(\det \gradq\x\right)^{-1}-1\right].
  \label{lagpoisson}
\end{eqnarray}
The density and peculiar velocity in Lagrangian variables are given by
\begin{equation}
  \begin{array}{r@{{}={}}l}
    \rho(\x(\q,\tau),\tau) & \left(\det \gradq\x\right)^{-1},\\[1ex]
    \v(\x(\q,\tau),\tau) & \Dtau\x(\q,\tau),
  \end{array}
\end{equation}
which automatically satisfy the mass conservation
law~\eqref{comovcontinuity}. Here $\Dtau$ is the operator of
Lagrangian time derivative, which in Lagrangian variables is the usual
partial time derivative at constant~$\q$ and in Eulerian variables
coincides with the material derivative $\dtau + \v\cdot\gradx$. The
notation~$\gradx$ in Lagrangian variables stands for the
$\x(\q,\tau)$-dependent differential operator with components
$\nabla_{x_{i}} \equiv
(\partial{q_{j}}/\partial{x_{i}})\nabla_{q_{j}}$, which expresses the
Eulerian gradient rewritten in Lagrangian coordinates, using the
inverse Jacobian matrix. Note that $\gradx$ and $\Dtau$ do not commute
and that terms with $\gradx$ in the Lagrangian equations are
implicitly non-linear.

In one dimension, \eqref{lagpoisson} has an interesting consequence:
\begin{equation}
  \label{prezeldovich}
  \nabla_{x} \phig  =-\frac{x-q}{\tau} .
\end{equation}
Indeed, in one dimension \eqref{lagpoisson} takes the form
\begin{equation}
  \label{lagpoisson1d}
  \nabla^2_x\phig = \frac{1}{\tau}\left[\left(\nabla_q x\right)^{-1}-1\right].
\end{equation}
Multiplying this equation by $\nabla_q x$ and expressing the first of
the two $x$-derivatives acting on~$\phig$ as a $q$-derivative, we
obtain
\begin{equation}
  \label{almostthere}
  \nabla_q\left(\nabla_x\phig\right) =\nabla_q\frac{q-x}{\tau}.
\end{equation}
Eq.~\eqref{prezeldovich} is obtained from \eqref{almostthere} by
integrating in~$q$. The absence of an arbitrary $\tau$-dependent
constant is established either by assuming vanishing at large
distances of both  $\phig$ and of the displacement $x-q$ or, in the
space-periodic case, by assuming the vanishing of period averages.
  
Using \eqref{prezeldovich} to eliminate the $\phig$ term in
\eqref{lagnewton} and introducing the notation~$\xi$ for the
displacement~$x-q$, we obtain
\begin{equation}
  \label{exact2ndorder}
  \DDtau\xi = -\frac{3}{2\tau}\left(\Dtau\xi-\frac{\xi}{\tau}\right).
\end{equation}
The only solution to this equation that remains well-behaved for $\tau
\to 0$ is the linear one $\xi \propto \tau$. This solution has the two
terms on the right-hand side of the one-dimensional version of
\eqref{lagnewton} cancelling each other and hence gives a vanishing
`acceleration' $\DDtau x$.

An approximate vanishing of acceleration takes place in higher
dimensions as well. For early times, the \emph{Lagrangian
map}~$\x(\q,\tau)$ stays close to the identity, with displacements
$\bxi(\q,\tau) = \x(\q,\tau) - \q$ small. Linearizing
\eqref{lagnewton} and~\eqref{lagpoisson} around  zero
displacement, we get the system
\begin{eqnarray}
  \label{linlagnewton}
  \DDtau\bxi &=& -\frac{3}{2\tau}(\Dtau\bxi + \gradq\phig),\\
  \label{linlagpoisson}
  \lapq\phig &=& -\frac1\tau \gradq\cdot\bxi.
\end{eqnarray}
Here we use the fact that $\gradx \simeq \gradq$ and $\det\gradq\x \simeq
\hbox{$1 + \gradq\cdot\bxi$}$. Using \eqref{linlagpoisson} to
eliminate~$\phig$ in~\eqref{linlagnewton}, we get for $\theta \equiv
\gradq\cdot\bxi$ an equation that coincides with~\eqref{exact2ndorder} up to
the change of variable $\xi \mapsto \theta$. Choosing the well-behaved linear
solution for~$\theta$, solving for~$\bxi$ and using the above argument to
eliminate a $\tau$-dependent constant, we see that, in the linearized
equations, terms in the right-hand side of \eqref{linlagnewton} cancel each
other and the acceleration vanishes. This simplification justifies using the
linear growth
factor $\tau$ as a time variable.

\section{History of mass transportation}
\setcounter{equation}{0}
\label{a:historymongeetc}

The subject of mass transportation was started by Gaspard Monge
\citeyearpar{M81} in a paper\footnote{The author's name appears in
this paper as `M.~Monge,' where the `M.' stands for `Monsieur.'} entitled
\textit{Th\'eorie des d\'eblais et des remblais} (Theory of cuts and
 fills) whose preamble is worth quoting entirely (our translation):
\begin{quote}
  $\quad$When earth is to be moved from one place to another, the usage is to
  call \emph{cuts} the volumes of earth to be transported and
  \emph{fills} the space to be occupied after transportation.

  $\quad$ The cost of transporting one molecule being, all things otherwise
  equal, proportional to its weight and to the distance
  [\textit{espace}] travelled and consequently the total cost being
  proportional to the sum of products of molecules each multiplied by
  the distance travelled, it follows that for given shapes and
  positions of the cuts and fills, it is not indifferent that any
  given molecule of the cuts be transported to this or that place in
  the fills, but there ought to be a certain distribution of
  molecules of the former into the latter, according to which the sum
  of these products will be the least possible, and the cost of
  transportation will be a {\it minimum.}
\end{quote}

Although clearly posed, the `mass transportation problem' was not solved, in
more than one dimension, until Leonid Kantorovich \citeyearpar{K42} formulated
a `relaxed' version, now called the Monge--Kantorovich problem: instead of a
`distribution of molecules of the former \emph{into} the latter,' he allowed a
distribution in the product space where more than one position in the fills
could be associated with a position in the cuts and where the initial and
final distributions are prescribed marginals (see Section~\ref{s:least}).  In
{\it cosmospeak}, he allowed multi-streaming with given initial and final mass
distributions. Using the techniques of duality and of linear programming that
he had invented (see Appendix~\ref{a:dualityproper}), Kantorovich was then
able to solve the mass transportation problem in this relaxed formulation.
The techniques developed by Kantorovich found many applications, notably in
economics, which in fact was his original motivation (he was awarded, together
with T.C.~Koopmans, the 1975 Nobel prize in this field).

Before turning to more recent developments we must say a few words
about the history of the Monge--Amp\`ere equation. 
It was considered for the first time by \citet{A20} for an unknown
function~$z(x,y)$ of two scalar variables. The equation is to be found
on p.~65 of Amp{\`e}re's huge (188 pages) mathematical memoir in the
form
\begin{equation}
  \label{amperehist}
  Hr + 2Ks + Lt + M + N(rt - s^2) = 0,
\end{equation}
where in modern notation $r = \partial^2 z/\partial x^2$, $s =
\partial^2 z/(\partial x\partial y)$, $t = \partial^2 z/\partial y^2$,
and $H, K, L, M, N$ are functions of $x, y, z$ and the two first-order
derivatives $p = \partial z/\partial x$ and $q=\partial z/\partial y$.
This extends the earlier work by \citet[see p.~126]{M84} concerning
the equation without the Hessian term ($N = 0$). Both Amp{\`e}re and
Monge were interested in methods of explicit integration of these
equations. Amp{\`e}re also pointed out the way the equation changes
under Legendre transformations but there is no physical interpretation
in terms of Lagrangian coordinates.\footnote{According to the
biography of Amp\`ere by L.~Pearce Williams in the {\it Dictionary of
Scientific Biography}, Amp\`ere's paper was written -- after he had
switched from mathematics to chemistry and physics -- with the purpose
of facilitating his election to the Paris Academy of Science; one can
then speculate that his mention of the Legendre transformation was
influenced by Legendre's presence in this academy.} There is evidence
that until the beginning of the 20th century the scientific community
attributed the equation with the Hessian solely to Amp\`ere
\citetext{\citet[p.~186]{B62} and \citet[p.~367]{EMW00}}. But the joint
attribution of \eqref{amperehist} to `Monge and Amp\`ere' is already
found in \citep{G96}.

The subjects of mass transportation and of the Monge--Amp\`ere
equation came together when one of us (YB) showed the equivalence of
the elliptic Monge--Amp\`ere equation and of the mass transportation
problem with quadratic cost: when initial and final distributions are
non-singular, the optimal solution is actually one-to-one, so that
nothing is lost by the Kantorovich relaxation trick
\citep{B87,B91}. For an extension of this result to general costs see
\citet{GM96}; a review of the many recent papers on the subject is
given by \citet{A03}.

\section{Basics of convexity and duality}
\label{a:duality}

\subsection{Convexity and the Legendre transformation}
\label{a:convexity}

A convex body may be defined by the condition that it coincides with
the intersection of all half-spaces containing it.  Obviously, it is
sufficient to take only those half-spaces limited by planes that touch
the body; such planes are called \emph{supporting}.

Take now a convex function $f(\q)$, so that the set of points in the
$(3+1)$-dimensional $(\q,f)$ space lying above its graph is convex. It follows
that we can write
\begin{equation}
  \label{fenchel}
  f(\q) = \max_\x\, \x\cdot\q - f^*(\x),
\end{equation}
where the expression $\x\cdot\q - f^*(\x)$ specifies a supporting
plane with the slope~$\x$ for the set of points lying above the graph
of~$f$ (see Fig.~\ref{f:legendre} for the one-dimensional case). 
\begin{figure}
  \iffigs
    \centerline{\psfig{file=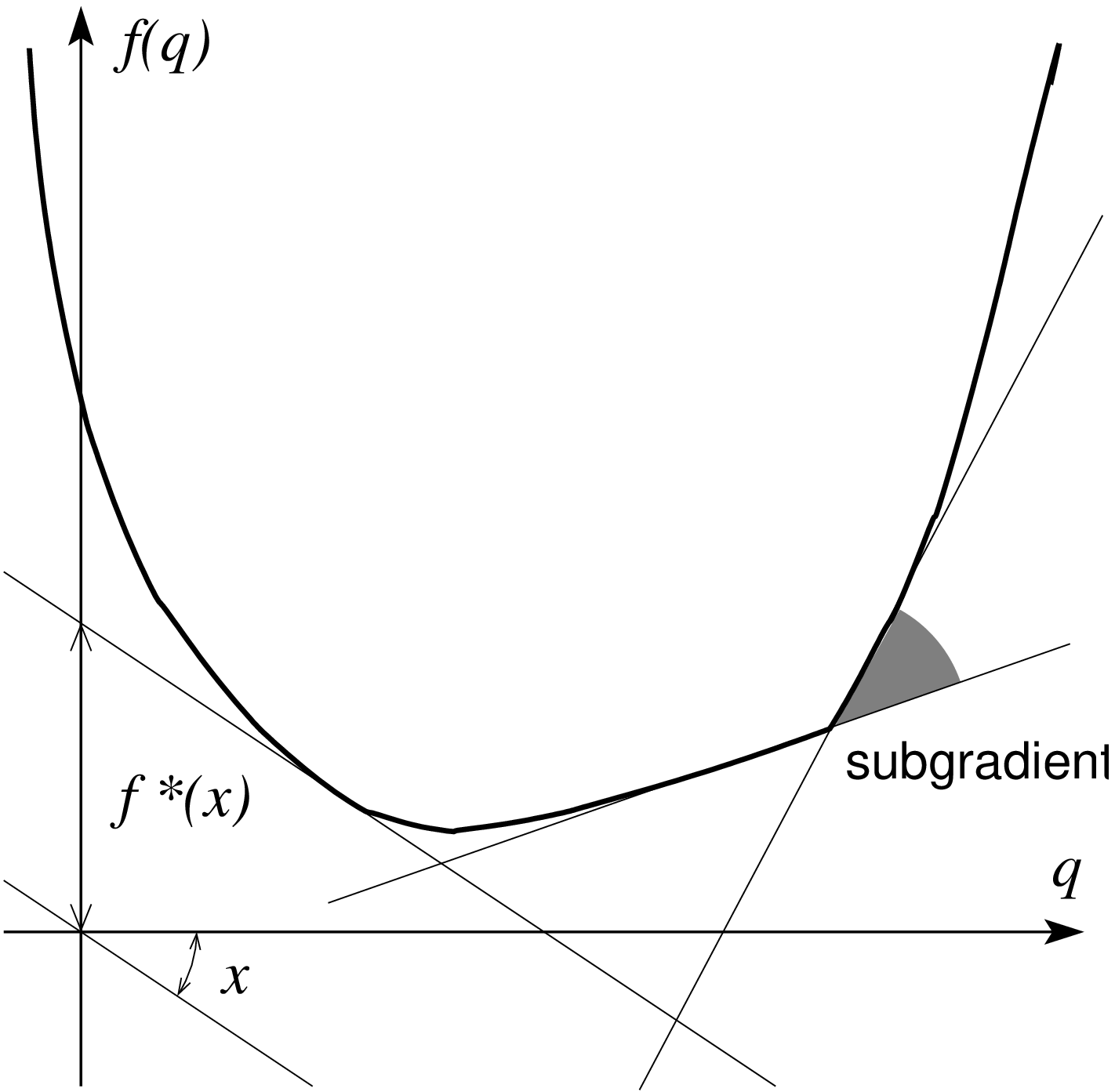,width=6.5cm}}
  \fi
  \vspace{2mm}
  \caption{A convex function $f(q)$ and the geometrical construction
    of its Legendre transform $f^*(x)$. Also illustrated is the
    subgradient of $f(q)$ at a non-smooth point.}
  \label{f:legendre}
\end{figure}
The function $f^*(\x)$, which specifies how high one should place a
supporting plane to touch the graph, is called the \emph{Legendre
transform} of~$f(\q)$.\footnote{It was introduced in the
one-dimensional case by \citet{M39} and then generalized by
\citet{F49}.}

 From Eq.~\eqref{fenchel} follows the inequality (known as the
\emph{Young inequality})
\begin{equation}
  \label{young}
  f(\q) + f^*(\x) \ge \x\cdot\q \mbox{ for all $\x$, $\q$,}
\end{equation}
where both sides coincide if and only if the supporting plane with the
slope~$\x$ touches the graph of~$f$ at~$\q$. This fact, together with
the obvious symmetry of this inequality, implies that
\begin{equation}
  \label{directfenchel}
  f^*(\x) = \max_\q\, \x\cdot\q - f(\q).
\end{equation}
Thus, the Legendre transform of a convex function is itself convex and
the Legendre transform of the Legendre transform recovers the initial
convex function.

If however we apply \eqref{fenchel} to a \emph{nonconvex}
 function~$f$, we obtain a convex function~$f^*$, whose own Legendre
transform will give the \emph{convex hull} of~$f$, the largest convex
 function whose graph lies below that of~$f$.

When $f$ is both convex and differentiable, \eqref{young} becomes an equality
for $\x = \gradq f(\q)$. If $f^*$ is also differentiable, then one also has
$\q = \gradx f^*(\x)$.  This is actually Legendre's original definition of the
transformation, which is thus limited to smooth functions. Furthermore, if the
original function is not convex and thus has the same gradient at separated
locations, Legendre's purely local definition will give a multivalued Legendre
transform.  (In the context of the present paper this corresponds to
multi-streaming.)

Not all convex functions are differentiable (e.g.\ $f(\q) = |\q|$).
But the Young inequality can be employed to define a useful
generalization of the gradient: the \emph{subgradient} of~$f$ at~$\q$
is the set of all~$\x$ for which the equality in~\eqref{young} holds
(see Fig.~\ref{f:legendre}).  If~$f$ is smooth at~$\q$, then $\gradq
 f(\q)$ will be the only such point; otherwise, there will be a
(convex) set of them.

If a convex function has the same subgradient at more than one point,
the function is said to lack \emph{strict convexity}.  In fact, strict
convexity and smoothness are complementary: lack of one in a convex
 function implies lack of the other in the Legendre transform.

 For further background on convex analysis and geometry, see
\citet{R70}.

\subsection{Duality in optimization}
\label{a:dualityproper}

Suppose we want to minimize a convex function $\Phi(\q)$ subject to a
set of linear constraints that may be written in matrix notation as
$A\q = \b$ (vectors $\q$ satisfying this constraint are called
\emph{admissible} in optimization parlance). We now observe that
\begin{equation}
  \label{prim0}
  \inf_{A\q = \b} \Phi(\q)
  = \myinf_{\q} \sup_{\x}\, \Phi(\q) - \x\cdot(A\q - \b).
\end{equation}
Indeed, should $A\q$ not equal~$\b$, the $\sup$ operation in~$\x$
will give infinity, so such $\q$ will not contribute to minimization.
Here we use the $\inf$/$\sup$ notation instead of $\min$/$\max$
because the extremal values may not be reached, e.g., when they are
infinite.

Using \eqref{fenchel}, we rewrite this in the form
\begin{equation}
  \label{dualprep0}
  \begin{array}{@{}l@{}}
    \displaystyle
    \myinf_{\q} \sup_{\x,\y}\,
    \y\cdot\q - \Phi^*(\y) - \x\cdot(A\q - \b) \\
    \displaystyle
    = \myinf_{\q} \sup_{\x,\y}\,
    (\y - A^T\x)\cdot\q - \Phi^*(\y) + \x\cdot\b,
  \end{array}
\end{equation}
where $\Phi^*(\y)$ is the Legendre transform of $\Phi(\q)$ and $A^T$
is the transpose of $A$. Taking $\inf$ in $\q$ first, we see that the
expression in the right-hand side will be infinite unless $\y =
A^T\x$. We then obtain the optimization problem of finding
\begin{equation}
  \label{dual0}
  \sup_{\x}\, \x\cdot\b - \Phi^*(A^T\x),
\end{equation}
which is called \emph{dual} to the original one. Note that there are
no constraints on the dual variable~$\x$: any value is admissible.

Denoting solutions of problems \eqref{prim0} and~\eqref{dual0} by
$\q^*$ and~$\x^*$, we see that
\begin{equation}
  \label{weakdual0}
  \Phi(\q^*) + \Phi^*(A^T\x^*) - \x^*\cdot\b = 0,
\end{equation}
because the optimal values of both problems are given
by~\eqref{dualprep0} and thus coincide. Furthermore, for any
admissible~$\q$ and~$\x$
\begin{equation}
  \label{weakdualineq}
  \Phi(\q) + \Phi^*(A^T\x) - \x\cdot\b \ge 0,
\end{equation}
because the right-hand sides of \eqref{prim0} and~\eqref{dual0} cannot
pass beyond their optimal values.

Moreover, let equality~\eqref{weakdual0} be satisfied for some
admissible $\q^*$ and $\x^*$; then such $\q^*$ and~$\x^*$ must solve
the problems \eqref{prim0} and~\eqref{dual0}. Indeed, taking e.g.\
$\x^*$ for~$\x$ in~\eqref{weakdualineq} and using~\eqref{weakdual0},
we see that for any other admissible~$\q$
\begin{equation}
  \Phi(\q^*) \le \Phi(\q),
  \label{aaa}
\end{equation}
i.e., that $\q^*$ solves the original optimization problem \eqref{prim0}.

Convex optimization problems with linear constraints considered in this
section are called \emph{convex programs}. Their close relatives are
\emph{linear programs}, namely optimization problems of the form
\begin{equation}
  \label{prim1} \inf_{A\q = \b,\, \q \ge 0} \c\cdot\q 
  = \myinf_{\q \ge 0} \sup_{\x}\, \c\cdot\q - \x\cdot(A\q - \b),
\end{equation}
where notation $\q \ge 0$ means that all components of the vector~$\q$
are nonnegative. Proceeding essentially as above with $\c\cdot\q$
instead of~$\Phi(\q)$, we observe that in order not to obtain infinity
when minimizing in~$\q$ in \eqref{dualprep0}, we have now to require
that $A^T\x \le \c$ (i.e.\ $\c - A^T\x \ge 0$). The dual problem thus
takes the form
\begin{equation}
  \label{dual1}
  \sup_{A^T\x \le \c}\, \x\cdot\b
\end{equation}
with an admissibility constraint on~$\x$. Instead of \eqref{weakdual0}
and~\eqref{weakdualineq} we obtain
\begin{equation}
  \label{weakdual1}
  \x^*\cdot\b = \c\cdot\q^* \quad \mbox{or} \quad (A^T\x^* - \c)\cdot\q^* = 0
\end{equation}
and
\begin{equation}
  \label{weakdualineq1}
  \x\cdot\b \le \c\cdot\q \quad \mbox{or} \quad(A^T\x - \c)\cdot\q \le 0,
\end{equation}
the latter inequality being automatically satisfied for any admissible
$\x,\q$. Note that for linear programs, the fact that
\eqref{weakdual1} holds for some admissible $\q^*,\x^*$ also implies
that $\q^*$ and~$\x^*$ solve their respective optimization problems.

 For further background on optimization and duality, see, e.g., \citet{PS82}.

\subsection{Why the analogue computer of Section~\protect\ref{s:nuts}
  solves the assignment problem}
\label{a:henon}

We suppose that the analogue computer described in
Section~\ref{s:nuts} has settled into equilibrium, which minimizes its
potential energy
\begin{equation}
  \label{potenrep}
  U = \sum_{i = 1}^N \alpha_i - \sum_{j = 1}^N \beta_j,
\end{equation}
under the set of constraints
\begin{equation}
  \label{discrdualconstrrep}
  \alpha_i - \beta_j \ge C-c_{ij},
\end{equation}
 for all $i,j$. Our goal is here is to show that the set of equilibrium
 forces $f_{ij}$, acting on studs between row and column rods, solves the
original linear programming problem of minimizing
\begin{equation}
  \label{relaxdiscractionrep}
  \tilde I = \sum_{i, j = 1}^N c_{ij} f_{ij}
\end{equation}
under constraints
\begin{equation}
  \label{discrconstraintsrep}
  f_{ij} \ge 0, \qquad \sum_{k = 1}^N f_{kj} = \sum_{k = 1}^N f_{ik} = 1,
\end{equation}
 for all $i,j$ and that in fact forces $f_{ij}$ take only zero and unit
values, thus providing the solution to the assignment problem.

Note first that if a row rod $A_i$ and a column rod $B_j$ are not in
contact at equilibrium, then the corresponding force vanishes ($f_{ij}
= 0$); if they are, then $f_{ij} \ge 0$.  Take now a particular pair
of rods $A_i$ and $B_j$ that are in contact. At equilibrium, the force
$f_{ij}$ must equal forces exerted on the corresponding stud by $A_i$
and $B_j$. We claim that both these forces must be integer. To see
this, let us compute the force exerted by $A_i$. This rod contributes
its weight, $+1$, possibly decreased by the force that it feels from
other column rods that are in contact with $A_i$. Each of these takes
$-1$ (its `buoyancy') out of the total force, but we may have to add
the force it feels in turn from other row rods with which it might be
in contact. Proceeding this way from one rod to another, we see that
all contributions, positive or negative, are unity, so their sum
$f_{ij}$ must be integer. The same argument applies to rod $B_j$.

Does this process indeed finish or, at some stage, do we come back at
an already visited stud and thus end up in an infinite cycle? In fact,
 for general set of stud lengths $C-c_{ij}$, the latter cannot happen,
because otherwise an alternating sum of some subset of stud lengths
would give exactly zero -- a zero probability event for a set of
arbitrary real numbers.

Consider now a row rod $A_i$. It is in contact with one or more column
rods, whose combined upward push must equilibrate the unit weight of
$A_i$.  Since any of the latter rods exerts a nonnegative integer
 force, it follows that exactly one of these forces is unity, and all
the other ones are zero. A similar argument holds for any column rod
$B_j$.

We have thus shown that all $f_{ij}$ in the equilibrium equal 1 or
0. One can of course ignore the vanishing forces.  Then each row rod
$A_i$ is supported by exactly one column rod $B_j$, and each $B_j$
supports exactly one $A_i$. This defines a one-to-one pairing, and we
are only left with a check that this pairing
minimizes~\eqref{relaxdiscractionrep}.

Observe that pushing a column rod down by some distance $\Delta$ and
simultaneously increasing by $\Delta$ the length of all studs attached
to this rod will have no effect on positions and constraints of all
other rods, hence on the equilibrium network of contacts. Moreover,
due to constraints \eqref{discrconstraintsrep}, the corresponding change
in coefficients $c_{ij}$ will not change the cost function
\eqref{relaxdiscractionrep} in any essential way, except of just
subtracting $\Delta$.

We can use this observation to put all column rods at the same level,
say at $z = 0$, adjusting $c_{ij}$ to some new values $c'_{ij}$.
Thus, for every $i$, the row rod $A_i$ rests on the stud with the
largest height $C-c'_{ij}$, so the equilibrium pairing maximizes the
sum
\begin{equation}
  \sum_{i, j = 1}^N (C - c'_{ij}) f_{ij}
\end{equation}
and thus minimizes \eqref{relaxdiscractionrep}.\footnote{Those readers
familiar with linear programming will recognize that the proof just presented
is based on two ideas: (i) the total unimodularity of the matrix of
constraints in terms of which the equalities in \eqref{discrconstraintsrep}
can be written and (ii) the complementary slackness \citep[see,
e.g.,][sections 3.2 and~13.2]{PS82}.}

\section{Details of the variational technique for the Euler--Poisson system}

\setcounter{equation}{0}
\label{a:details}

In this appendix, we explain details of the variational procedure
outlined in Section~\ref{s:selfgravitating}, which proves that
prescription of the density fields at terminal epochs $\tau = 0$ and
$\tau = \tau_0$ uniquely determines a regular and thus curl-free
solution to the Euler--Poisson system
\eqref{comoveuler}--\eqref{comovpoisson}.

The variational problem is posed for the functional
\begin{equation}
  \label{actionrep}
  I = \frac12 \int_0^{\tau_0}\d\tau \int\D\x\, \tau^{3/2} 
  \left(\rho|\v|^2 + \frac32|\gradx\phig|^2\right)
\end{equation}
with four constraints: the Poisson equation \eqref{comovpoisson}, which we
repeat here for convenience,
\begin{equation}
  \label{comovpoissonrep}
  \lapx\phig = \frac{\rho - 1}{\tau},
\end{equation}
the mass conservation \eqref{comovcontinuity}, also repeated here,
\begin{equation}
  \label{comovcontinuityrep}
  \dtau\rho + \gradx\cdot(\rho\v) = 0,
\end{equation}
and the two boundary conditions
\begin{equation}
  \label{boundary}
  \rho(\x, 0) = 1 \quad {\rm and} \quad \rho(\x, \tau_0) = \rho_0(\x).
\end{equation}
In the sequel, we shall always denote by $\iint$ the double integration
over $0\le\tau\le\tau_0$ and over the whole space domain in $\x$
provided that the integrand vanishes at infinity sufficiently fast, or
over the periodicity box in the case of periodic boundary conditions.
A single integral sign $\int$ will always denote the integration over
the relevant space domain in $\x$.

 First, we make this problem convex by rewriting the functional and
constraints in a new set of variables with the mass flux $\J(\x,t) =
\rho(\x,t)\,\v(\x,t)$ instead of the velocity~$\v$. The mass conservation
constraint, which was the only non-linear one in the old variables,
becomes now linear:
\begin{equation}
  \label{contj}
  \dtau\rho + \gradx\cdot\J = 0,
\end{equation}
and one can check that the density of kinetic energy takes the form
\[
  \frac12\rho|\v|^2 =
  \frac{1}{2\rho}|\J|^2
  = \max_{c,\, \m\colon\,\, c + |\m|^2/2 \le 0}\, (\rho c + \J\cdot\m)
\]
or 
\begin{equation}
  \label{temp0}
  \frac{|\J|^2}{2\rho} = \max_{c,\, \m}\, (\rho c + \J\cdot\m - F(c,\m)),
\end{equation}
where
\begin{equation}
  F(c,\m) = \cases{0 & if $c + |\m|^2/2 \le 0$\cr +\infty & otherwise.\cr}
\end{equation}
Note that in \eqref{temp0} the variables $c,\m$, as well as $\rho,
\J$, are functions of $(\x,\tau)$. The action functional may now be
written as
\begin{equation}
  I = \frac12\iint \left(\frac{1}{\rho}|\J|^2 + \frac32|\gradx\phi|^2\right)
  \tau^{3/2}\, \D\x\, \d\tau,
\end{equation}
and turns out to be convex.

To see this, first note that the operation of integration is linear
and thus preserves convexity of the integrand. The integrand is a
positive quadratic function of $\gradx\phi$ and therefore is convex in
$\phi$; furthermore,  \eqref{temp0} implies that it is also
convex in $(\rho,\J)$, since the kinetic energy density $|\J|^2/2\rho$
is the Legendre transform of the function $F(c,\m)$, which itself is
convex. 

Note also that by representing the kinetic energy density in the form
\eqref{temp0}, we may safely allow $\rho$ to take negative values: the
right-hand side being in that case $+\infty$, it will not contribute to
minimizing \eqref{actionrep}. 

We now  derive the dual optimization problem. We
introduce the scalar Lagrange multipliers $\psi(\x,t)$,
$\vartheta_{\mathrm{in}}(\x)$, $\vartheta_0(\x)$ and $\theta(\x,t)$
 for the Poisson equation \eqref{comovpoissonrep}, the boundary
conditions \eqref{boundary}, and the constraints of mass conservation
\eqref{contj}, respectively, and observe that the variational problem
may now be written in the form
\begin{equation}
  \label{saddle1}
  \begin{array}{@{}l@{}}
    \displaystyle
    \inf_{\begin{array}{@{}c@{}}
      \rule{0pt}{10.5pt}\scriptstyle\rho,\J,\phi\\
    \end{array}}\;
    \sup_{\begin{array}{@{}c@{}}
      \scriptstyle c, \m, \theta, \psi,\vartheta_0,\vartheta_T\colon\\
      \scriptstyle c + |\m|^2/2 \le 0
    \end{array}}
    \iint\D\x\,\d\tau \left[
    \frac32\psi\left(\lapx\phi-\frac{\rho-1}{\tau}\right)\right.\\
    \displaystyle\quad\left. 
    +\theta(\dtau\rho + \gradx\cdot\J) +
    \tau^{3/2}\left(\rho c+\J\cdot\m+\frac34|\gradx\phi|^2\right)\right]\quad\\
    \displaystyle\quad
    +\int\vartheta_{\mathrm{in}}(\x)(\rho(\x,0)-1)\, \D\x\\ 
    \displaystyle\quad
    -\int\vartheta_0(\x) (\rho(\x,\tau_0)-\rho_0(\x))\, \D\x.
  \end{array}
\end{equation}
To see that \eqref{saddle1} is indeed equivalent to minimizing
\eqref{actionrep} under the constraints \eqref{comovcontinuityrep} or
\eqref{contj}, \eqref{comovpoissonrep}, and \eqref{boundary}, observe
that for those $\rho,\J,\phi$ that do not satisfy the constraints, the
$\sup$ operation over $\theta, \psi, \vartheta_{\mathrm{in}},
\vartheta_0$ will give positive infinity; the $\sup$ will be finite
(and thus contribute to the subsequent minimization) only if all
constraints are satisfied.  (This argument is the functional version
of what is explained in Appendix~\ref{a:dualityproper} for the
 finite-dimensional case.)

Performing an integration by parts in the $\tau$ variable in
\eqref{saddle1} and using the boundary conditions on the mass density
\eqref{boundary}, we find that $\vartheta_{\mathrm{in}}(\x) =
\theta(\x,0)$ and $\vartheta_0(\x) = \theta(\x,\tau_0)$. Integrating
 further by parts in the $\x$ variable, assuming that boundary terms at
infinity vanish (or that we have periodic boundary conditions in
space) and rearranging terms, we get
\begin{equation}
\begin{array}{@{}l@{}}
  \displaystyle
  \inf_{\begin{array}{@{}c@{}}
    \rule{0pt}{10.5pt}\scriptstyle\rho,\J,\phi\\
  \end{array}}\!
  \sup_{\begin{array}{@{}c@{}}
    \scriptstyle c, \m, \theta, \psi\colon\\[-0.5ex]
    \scriptstyle c + |\m|^2/2 \le 0
  \end{array}}
  \iint\D\x\, \d\tau 
  \left(\rho\,(c\tau^{3/2} - \dtau\theta - \frac3{2\tau}\psi)\right. \\
  \displaystyle\qquad
  + \J\cdot(\m\tau^{3/2} - \gradx\theta)
  + \frac3{4\tau^{3/2}}\,|\gradx\psi - \tau^{3/2}\gradx\phig|^2 \\[2ex]
  \displaystyle\qquad\left.
  - \frac3{4\tau^{3/2}}\,|\gradx\psi|^2 + \frac3{2\tau}\psi\right)\, \\[2ex]
  \displaystyle\qquad 
  - \int \theta(\x,0)\, \D\x 
    + \int \theta(\x,\tau_0)\, \rho_0(\x)\, \D\x.
\end{array}
\label{dualprep}
\end{equation}
Performing minimization with respect to $\rho,\J,\phi$ first, as in
\eqref{dualprep0} of Appendix~\ref{a:dualityproper}, we see that the
 following two equalities must hold (remember that $\rho$ need not be
positive at this stage):
\begin{equation}
  c = \frac1{\tau^{3/2}}\left(\dtau\theta + \frac{3\psi}{2\tau}\right),\qquad 
  \m = \frac1{\tau^{3/2}}\gradx\theta,
  \label{cond1}
\end{equation}
so that terms linear in $\rho$ and~$\J$ vanish in~\eqref{dualprep}. It
 follows that $c$ and $\m$ are determined by $\theta$ and $\psi$ and
that the constraint $c + |\m|^2/2 \le 0$ can be written
\begin{equation}
  \label{dualconstrrep} 
  \dtau\theta + \frac1{2\tau^{3/2}}|\gradx\theta|^2 + \frac3{2\tau}\psi \le 0.
\end{equation}
Also, the $\inf$ with respect to $\phi$ is straightforward and gives
\begin{equation}
  \tau^{3/2}\gradx\phig = \gradx\psi. 
  \label{cond2}
\end{equation}
Using \eqref{cond1} and~\eqref{cond2} in~\eqref{dualprep}, we arrive
at the optimization problem of maximizing
\begin{equation}
  \label{actiondualrep}
  \begin{array}{r@{}l}
    J&\displaystyle{} 
    = \iint\left(\frac3{2\tau}\psi-\frac3{4\tau^{3/2}}|\gradx\psi|^2\right)\,
    \D\x\, \d\tau\\[3ex]
    &\displaystyle + \int \theta(\x,\tau_0)\, \rho_0(\x)\, \D\x
    - \int \theta(\x,0)\, \D\x
  \end{array}
\end{equation}
under constraint \eqref{dualconstrrep}. Eqs.\ \eqref{actiondualrep}
and~\eqref{dualconstrrep} constitute a variational problem \emph{dual}
to the original one.

As both the original and the dual variational problems have the same
saddle-point formulation \eqref{saddle1} or~\eqref{dualprep}, the
optimal values of the two functionals \eqref{actionrep}
and~\eqref{actiondualrep} are equal. Let $(\rho,\J,\phig)$ be a
solution to the original variational problem and $\theta,\psi$ be a
solution to the dual one. Subtracting the (equal) optimal values from
each other, we may now write, similarly to \eqref{weakdual0},
\begin{equation}
  \label{weakdual}
  \begin{array}{@{}l@{}}
    \displaystyle\iint
    \biggl(\frac{\tau^{3/2}}{2\rho}|\J|^2 
    +\frac{3\tau^{3/2}}{4}|\gradx\phig|^2\\[3ex]
    \displaystyle\qquad
    +\frac3{4\tau^{3/2}}|\gradx\psi|^2 - \frac3{2\tau}\psi\biggr)\, 
    \D\x\,\d\tau\\[2ex]
    \displaystyle\qquad
    + \int \theta(\x,0)\, \D\x 
    -\int \theta(\x,\tau_0)\, \rho_0(\x)\, \D\x = 0.
  \end{array}
\end{equation}
We are going to show that the left-hand side of \eqref{weakdual} may be given
the form of a sum of three nonnegative terms, each of which will
therefore have to vanish. First, we rewrite the last two integrals,
using the mass conservation constraint \eqref{contj} and integrations
by parts, in the form
\[
  -\iint \dtau(\theta\rho)\, \D\x\,\d\tau
  = -\iint
  (\dtau\theta\,\rho + \gradx\theta\cdot\J)\, \D\x\,\d\tau.
\]
Second, we note that
\[
\begin{array}{@{}l@{}}
  \displaystyle 
  \iint
  \left(\frac{3\tau^{3/2}}{4}|\gradx\phig|^2 + 
  \frac3{4\tau^{3/2}}|\gradx\psi|^2\right)\, \D\x\,\d\tau\\[3ex]
  \displaystyle 
  = \iint
  \left(\frac3{4\tau^{3/2}}|\tau^{3/2}\gradx\phig - \gradx\psi)|^2
    - \frac3{2\tau}\psi(\rho - 1)\right)\, \D\x\,\d\tau, \end{array}
\] 
which follows from the Poisson constraint \eqref{comovpoissonrep}.
Taking all this into account in \eqref{weakdual}, we get, after a
rearrangement of terms, 
\begin{equation}
  \label{compslackness}
  \begin{array}{l}
    \displaystyle
    \iint \frac{\rho}{2\tau^{3/2}}
    \left|\frac{\tau^{3/2}}{\rho}\J-\gradx\theta\right|^2\, \D\x\,\d\tau\\[4ex]
    \displaystyle\quad
    +\iint-\rho\left(\dtau\theta+\frac1{2\tau^{3/2}}|\gradx\theta|^2
    +\frac3{2\tau}\psi\right)\, \D\x\,\d\tau \\[4ex]
    \displaystyle\quad
    +\iint\frac3{4\tau^{3/2}}
    |\tau^{3/2}\gradx\phig-\gradx\psi)|^2\, \D\x\,\d\tau=0.
  \end{array}
\end{equation}
The left-hand side is a sum of three nonnegative terms (the second is so
by~\eqref{dualconstrrep}), all of which must thus vanish. This gives
\begin{equation}
  \label{compslackness1}
  \v = \frac1\rho\J = \frac1{\tau^{3/2}}\gradx\theta,\qquad
  \gradx\phig = \frac1{\tau^{3/2}}\gradx\psi
\end{equation}
and
\begin{equation}
  \label{compslackness2}
  \dtau\theta + \frac1{2\tau^{3/2}}|\gradx\theta|^2 
  + \frac3{2\tau}\psi = 0,
\end{equation}
wherever $\rho$ is non-vanishing (otherwise the left-hand-side is
non-positive by \eqref{dualconstrrep}). The last equality turns into
the Euler equation
\begin{equation}
  \label{comoveulerrep}
  \dtau\v + (\v\cdot\gradx)\v = -\frac3{2\tau}(\v + \gradx\phig)
\end{equation}
by taking the gradient and using \eqref{compslackness1}. 

By \eqref{compslackness1} and~\eqref{compslackness2}, any two hypothetically
different minimizing solutions for either variational problem give
rise to the same velocity potential and to the same gravitational potential (up
to insignificant constants)  and thus define the same solution $(\rho,
\v, \phig)$ to the Euler--Poisson equations with the  boundary
conditions~\eqref{boundary} and the condition of curl-free velocity.

Moreover, for any such solution $(\rho, \v, \phig)$, one can
use~\eqref{compslackness1} to define $\theta$ and~$\psi$ that
satisfy~\eqref{compslackness2} and thus~\eqref{dualconstrrep}. By
\eqref{compslackness}, the values of functionals $I$ and~$\bar I$ evaluated at
these functions will coincide; together with convexity this implies, by an
argument similar to that given in Appendix~\ref{a:dualityproper} concerning
\eqref{aaa}, that such $(\rho, \v, \phig)$ and $(\theta, \psi)$ in fact
minimize both functionals under the corresponding constraints.

This means that a (curl-free) velocity field, a gravitational field and a
density fields $(\v, \phig,\rho)$ will satisfy the Euler--Poisson equations
\eqref{comoveuler}--\eqref{comovpoisson} (repeated as 
\eqref{comoveulerrep}, \eqref{comovcontinuityrep}, and \eqref{comovpoissonrep}
in this Appendix) and the boundary conditions~\eqref{boundary} if and only if
they minimize \eqref{actionrep} under the corresponding constraints.  This
establishes uniqueness.

\endgroup



\begin{thebibliography}{}

\bibitem[\protect\citeauthoryear{Ambrosio}{Ambrosio}{2003}]{A03}
Ambrosio L.,  2003, in Proceedings of the ICM Beijing 2002, Vol. 3,
pp 131--140 and arXiv.org/abs/math.AP/0304389

\bibitem[\protect\citeauthoryear{{Amp{\`e}re}}{{Amp{\`e}re}}{1820}]{A20}
{Amp{\`e}re} A.-M.,  1820, J. de L'\'Ecole Royale Polytechnique, 11, 1

\bibitem[\protect\citeauthoryear{{Arnol'd}, {Shandarin} \&
  {Zel'dovich}}{{Arnol'd} et~al.}{1982}]{AZS82}
{Arnol'd} V.~I.,  {Shandarin} S.~F.,    {Zel'dovich} Y.~B.,  1982, Geophys.
  Astrophys. Fluid Dynamics, 20, 111

\bibitem[\protect\citeauthoryear{Balinski}{Balinski}{1986}]{B86}
Balinski M.~L.,  1986, Math. Programming, 34, 125

\bibitem[\protect\citeauthoryear{Bec, Frisch \& Khanin}{Bec
  et~al.}{2000}]{BFK00}
Bec J.,  Frisch U.,    Khanin K.,  2000, J. Fluid Mech., 416, 239

\bibitem[\protect\citeauthoryear{{Benamou} \& {Brenier}}{{Benamou} \&
  {Brenier}}{2000}]{BB00}
{Benamou} J.-D.,  {Brenier} Y.,  2000, Numer. Math., 84, 375

\bibitem[\protect\citeauthoryear{Bernardeau, Colombi, Gazta{\~n}aga \&
  Scoccimarro}{Bernardeau et~al.}{2002}]{BSGS02}
Bernardeau F.,  Colombi S.,  Gazta{\~n}aga E.,    Scoccimarro R.,  2002, Phys.
  Rep., 367, 1

\bibitem[\protect\citeauthoryear{{Bertschinger} \& {Dekel}}{{Bertschinger} \&
  {Dekel}}{1989}]{BD89}
{Bertschinger} E.,  {Dekel} A.,  1989, ApJ, 336, L5

\bibitem[\protect\citeauthoryear{Bertsekas}{Bertsekas}{1981}]{B81}
Bertsekas D.~P.,  1981, Math. Programming, 21, 152

\bibitem[\protect\citeauthoryear{Bertsekas}{Bertsekas}{1992}]{B92a}
Bertsekas D.~P.,  1992, Comput. Optim. Appl., 1, 7

\bibitem[\protect\citeauthoryear{Bertsekas}{Bertsekas}{2001}]{B01}
Bertsekas D.,  2001, in Encyclopedia of optimization, Vol.~I.
Kluwer Academic Publishers, Dordrecht

\bibitem[\protect\citeauthoryear{Bour}{Bour}{1862}]{B62}
Bour {\'E}.,  1862, Journal de l'{\'E}cole polytechnique, 22, cah.~39, 149

\bibitem[\protect\citeauthoryear{{Branchini}, {Eldar} \& {Nusser}}{{Branchini}
  et~al.}{2002}]{BEN02}
{Branchini} E.,  {Eldar} A.,    {Nusser} A.,  2002, MNRAS, 335, 53

\bibitem[\protect\citeauthoryear{{Brenier}}{{Brenier}}{1987}]{B87}
{Brenier} Y.,  1987, C. R. Acad. Sci. Paris S\'er. I Math., 305, 805

\bibitem[\protect\citeauthoryear{{Brenier}}{{Brenier}}{1991}]{B91}
{Brenier} Y.,  1991, Comm. Pure Appl. Math., 44, 375

\bibitem[\protect\citeauthoryear{Bridle, Lahav, Ostriker \& Steinhardt}{Bridle
  et~al.}{2003}]{BLOS03}
Bridle S.~L.,  Lahav O.,  Ostriker J.~P.,    Steinhardt P.~J.,  2003, Sci,
  299, 1532

\bibitem[\protect\citeauthoryear{{Buchert}}{{Buchert}}{1992}]{B92}
{Buchert} T.,  1992, MNRAS, 254, 729

\bibitem[\protect\citeauthoryear{{Buchert} \& {Dominguez}}{{Buchert} \&
  {Dominguez}}{1998}]{BD98}
{Buchert} T.,  {Dominguez} A.,  1998, A\&A, 335, 395

\bibitem[\protect\citeauthoryear{{Buchert} \& {Ehlers}}{{Buchert} \&
  {Ehlers}}{1993}]{BE93}
{Buchert} T.,  {Ehlers} J.,  1993, MNRAS, 264, 375

\bibitem[\protect\citeauthoryear{Burkard \& Derigs}{Burkard \&
  Derigs}{1980}]{BD80}
Burkard R.~E.,  Derigs U.,  1980, Assignment and matching problems: solution
  methods with {FORTRAN} programs.
Vol.~184 of Lecture Notes in Economics and Mathematical Systems,
  Springer-Verlag, Berlin

\bibitem[\protect\citeauthoryear{Caffarelli \& Li}{Caffarelli \&
  Li}{2001}]{CL01}
Caffarelli L.,  Li Y.~Y.,  2001, An extension to a theorem of {J\"o}rgens,
  {C}alabi, and {P}ogorelov, preprint ({\small
  www.math.utexas.edu/users/combs/Caffarelli/extension.pdf})

\bibitem[\protect\citeauthoryear{Caffarelli}{Caffarelli}{1999}]{C99}
Caffarelli L.~A.,  1999, in Christ M.,  Kenig C.~E.,   Sadosky C.,  eds,
  Chicago Lectures in Math., Harmonic analysis and partial differential
  equations.
Univ. Chicago Press, Chicago, IL, pp 117--126

\bibitem[\protect\citeauthoryear{Caffarelli \& Milman}{Caffarelli \&
  Milman}{1999}]{MAE99}
Caffarelli L.~A.,  Milman M.,  eds, 1999, Monge {A}mp\`ere equation:
  applications to geometry and optimization.
Vol.~226 of Contemporary Mathematics, American Mathematical Society,
  Providence, RI

\bibitem[\protect\citeauthoryear{{Catelan}}{{Catelan}}{1995}]{C95}
{Catelan} P.,  1995, MNRAS, 276, 115

\bibitem[\protect\citeauthoryear{{Catelan}, {Lucchin}, {Matarrese} \&
  {Moscardini}}{{Catelan} et~al.}{1995}]{CLMM95}
{Catelan} P.,  {Lucchin} F.,  {Matarrese} S.,    {Moscardini} L.,  1995, MNRAS,
  276, 39

\bibitem[\protect\citeauthoryear{{Coles} \& {Lucchin}}{{Coles} \&
  {Lucchin}}{2002}]{CL95}
{Coles} P.,  {Lucchin} F.,  2002, {Cosmology: The Origin and Evolution of
  Cosmic Structure}.
John Wiley \& Sons, Chichester et al.

\bibitem[\protect\citeauthoryear{{Couchman}, {Thomas} \& {Pearce}}{{Couchman}
  et~al.}{1995}]{CTP95}
{Couchman} H.~M.~P.,  {Thomas} P.~A.,    {Pearce} F.~R.,  1995, ApJ, 452, 797

\bibitem[\protect\citeauthoryear{Croft \& {Gazta{\~{n}}aga}}{Croft \&
  {Gazta{\~{n}}aga}}{1997}]{CG97}
Croft R.~A.~C.,  {Gazta{\~{n}}aga} E.,  1997, MNRAS, 285, 793

\bibitem[\protect\citeauthoryear{{Dekel}, {Bertschinger} \& {Faber}}{{Dekel}
  et~al.}{1990}]{DBF90}
{Dekel} A.,  {Bertschinger} E.,    {Faber} S.~M.,  1990, ApJ, 364, 349

\bibitem[\protect\citeauthoryear{{Fanelli} \& {Aurell}}{{Fanelli} \&
  {Aurell}}{2002}]{FA02}
{Fanelli} D.,  {Aurell} E.,  2002, A\&A, 395, 399

\bibitem[\protect\citeauthoryear{Fenchel}{Fenchel}{1949}]{F49}
 Fenchel W.,  1949, Canadian J. Math., 1, 73

\bibitem[\protect\citeauthoryear{Fr{\'e}chet}{Fr{\'e}chet}{1957a}]{F57a}
 Fr{\'e}chet M.,  1957a, C. R. Acad. Sci. Paris S\'er. I Math., 244, 689

\bibitem[\protect\citeauthoryear{Fr{\'e}chet}{Fr{\'e}chet}{1957b}]{F57b}
 Fr{\'e}chet M.,  1957b, Publ. Inst. Statist. Univ. Paris, 6, 183

\bibitem[\protect\citeauthoryear{{Frieman} \& {Szalay}}{{Frieman} \&
  {Szalay}}{2000}]{FS00}
{Frieman} J.~A.,  {Szalay} A.~S.,  2000, Phys. Rep., 333, 215

\bibitem[\protect\citeauthoryear{Frisch \& Bec}{Frisch \& Bec}{2002}]{FB02}
 Frisch U.,  Bec J.,  2002, in Lesieur M.,  Yaglom A.,   David F.,  eds,
  {\'E}cole de physique des {H}ouches, session {LXXIV}, New trends in
  turbulence.
EDP Sciences and Springer, pp 341--384

\bibitem[\protect\citeauthoryear{{Frisch}, {Matarrese}, {Mohayaee} \&
  {Sobolevski}}{{Frisch} et~al.}{2002}]{FMMS02}
{Frisch} U.,  {Matarrese} S.,  {Mohayaee} R.,    {Sobolevski} A.,  2002, Nat,
  417, 260

\bibitem[\protect\citeauthoryear{{Gangbo} \& {McCann}}{{Gangbo} \&
  {McCann}}{1996}]{GM96}
{Gangbo} W.,  {McCann} R.~J.,  1996, Acta Math., 177, 113

\bibitem[\protect\citeauthoryear{{Giavalisco}, {Mancinelli}, {Mancinelli} \&
  {Yahil}}{{Giavalisco} et~al.}{1993}]{GMMY93}
{Giavalisco} M.,  {Mancinelli} B.,  {Mancinelli} P.~J.,    {Yahil} A.,  1993,
  ApJ, 411, 9

\bibitem[\protect\citeauthoryear{Goursat}{Goursat}{1896}]{G96}
Goursat {\'E}.,  1896, Le{\c c}ons sur l'int{\'e}gration des {\'e}quations aux
  d{\'e}riv{\'e}es partielles du second ordre.
Vol.~I, Paris

\bibitem[\protect\citeauthoryear{{Gramann}}{{Gramann}}{1993}]{G93}
{Gramann} M.,  1993, ApJ, 405, 449

\bibitem[\protect\citeauthoryear{{Gurbatov} \& {Saichev}}{{Gurbatov} \&
  {Saichev}}{1984}]{GS84}
{Gurbatov} S.~N.,  {Saichev} A.~I.,  1984, Izv. Vys{\v{s}}. U{\v{c}}ebn. Zaved.
  Radiofizika, 27, 456

\bibitem[\protect\citeauthoryear{{Gurbatov}, {Saichev} \&
  {Shandarin}}{{Gurbatov} et~al.}{1989}]{GSS89}
{Gurbatov} S.~N.,  {Saichev} A.~I.,    {Shandarin} S.~F.,  1989, MNRAS, 236,
  385

\bibitem[\protect\citeauthoryear{H{\'e}non}{H{\'e}non}{1995}]{H95}
H{\'e}non M.,  1995, C. R. Acad. Sci. Paris S\'er. I Math., 321, 741

\bibitem[\protect\citeauthoryear{H{\'e}non}{H{\'e}non}{2002}]{H02}
H{\'e}non M., 2002, A mechanical model for the transportation problem,
  preprint (arXiv:math.OC/0209047)

\bibitem[\protect\citeauthoryear{{Jeans}}{{Jeans}}{1919}]{J19}
{Jeans} J.~H.,  1919, {Problems of Cosmogony and Stellar Dynamics}.
Cambridge University Press, Cambridge

\bibitem[\protect\citeauthoryear{{Kantorovich}}{{Kantorovich}}{1942}]{K42}
{Kantorovich} L.~V.,  1942, C. R. (Doklady) Acad. Sci. USSR, 321, 199

\bibitem[\protect\citeauthoryear{{Kardar}, {Parisi} \& {Zhang}}{{Kardar}
  et~al.}{1986}]{KPZ86}
{Kardar} M.,  {Parisi} G.,    {Zhang} Y.,  1986, Phys. Rev. Lett., 56, 889

\bibitem[\protect\citeauthoryear{{Kolatt}, {Dekel}, {Ganon} \&
  {Willick}}{{Kolatt} et~al.}{1996}]{KDGW96}
{Kolatt} T.,  {Dekel} A.,  {Ganon} G.,    {Willick} J.~A.,  1996, ApJ, 458, 419

\bibitem[\protect\citeauthoryear{Kuhn}{Kuhn}{1955}]{K55}
Kuhn H.~W.,  1955, Naval Res. Logist. Quart., 2, 83

\bibitem[\protect\citeauthoryear{Landau \& Lifshitz}{Landau \&
  Lifshitz}{1960}]{LL60}
Landau L.~D.,  Lifshitz E.~M.,  1960, Mechanics.
Vol.~1 of Course of Theoretical Physics, Pergamon Press, Oxford

\bibitem[\protect\citeauthoryear{Loeper}{Loeper}{2003}]{L03}
Loeper G.,  2003, The inverse problem for the {Euler--Poisson} system in
  cosmology, preprint (math.AP/0306430)

\bibitem[\protect\citeauthoryear{Mandelbrojt}{Mandelbrojt}{1939}]{M39}
Mandelbrojt S.,  1939, C. R. Acad. Sci. Paris, 209, 977

\bibitem[\protect\citeauthoryear{Minkowski}{Minkowski}{1897}]{M97}
Minkowski H.,  1897, {N}achr. {G}es. {W}iss. {G\"o}ttingen ({M}ath. {P}hys.
  {K}lasse), pp 198--219

\bibitem[\protect\citeauthoryear{Mohayaee, Frisch, Matarrese \&
  Sobolevski{\u\i}}{Mohayaee et~al.}{2003}]{MFMS03}
Mohayaee R.,  Frisch U.,  Matarrese S., Sobolevski{\u\i} A., 2003, A\&A, 406,
  393--401

\bibitem[\protect\citeauthoryear{{Monaco} \& {Efstathiou}}{{Monaco} \&
  {Efstathiou}}{1999}]{ME99}
{Monaco} P.,  {Efstathiou} G.,  1999, MNRAS, 308, 763

\bibitem[\protect\citeauthoryear{Monge}{Monge}{1781}]{M81}
Monge G.,  1781, Histoire de l'Acad{\'e}mie Royale des Sciences, pp 666--704

\bibitem[\protect\citeauthoryear{Monge}{Monge}{1784}]{M84}
Monge G.,  1784, Histoire de l'Acad{\'e}mie Royale des Sciences, pp 118--192

\bibitem[\protect\citeauthoryear{{Moutarde}, {Alimi}, {Bouchet}, {Pellat} \&
  {Ramani}}{{Moutarde} et~al.}{1991}]{MABPR91}
{Moutarde} F.,  {Alimi} J.-M.,  {Bouchet} F.~R.,  {Pellat} R.,    {Ramani} A.,
  1991, ApJ, 382, 377

\bibitem[\protect\citeauthoryear{{Munshi}, {Sahni} \& {Starobinsky}}{{Munshi}
  et~al.}{1994}]{MSS94}
{Munshi} D.,  {Sahni} V.,    {Starobinsky} A.~A.,  1994, ApJ, 436, 517

\bibitem[\protect\citeauthoryear{{Narayanan} \& {Croft}}{{Narayanan} \&
  {Croft}}{1999}]{NC99}
{Narayanan} V.~K.,  {Croft} R.~A.~C.,  1999, ApJ, 515, 471

\bibitem[\protect\citeauthoryear{{Narayanan} \& {Weinberg}}{{Narayanan} \&
  {Weinberg}}{1998}]{NW98}
{Narayanan} V.~K.,  {Weinberg} D.~H.,  1998, ApJ, 508, 440

\bibitem[\protect\citeauthoryear{{Nusser} \& {Branchini}}{{Nusser} \&
  {Branchini}}{2000}]{NB00}
{Nusser} A.,  {Branchini} E.,  2000, MNRAS, 313, 587

\bibitem[\protect\citeauthoryear{{Nusser} \& {Dekel}}{{Nusser} \&
  {Dekel}}{1992}]{ND92}
{Nusser} A.,  {Dekel} A.,  1992, ApJ, 391, 443

\bibitem[\protect\citeauthoryear{Papadimitriou \& Steiglitz}{Papadimitriou \&
  Steiglitz}{1982}]{PS82}
Papadimitriou C.~H.,  Steiglitz K.,  1982, Combinatorial optimization:
  algorithms and complexity.
Prentice-Hall Inc., Englewood Cliffs, NJ

\bibitem[\protect\citeauthoryear{{Peebles}}{{Peebles}}{1980}]{P80}
{Peebles} P.~J.~E.,  1980, {The large-scale structure of the universe}.
Princeton University Press, Princeton, NJ

\bibitem[\protect\citeauthoryear{{Peebles}}{{Peebles}}{1989}]{P89}
{Peebles} P.~J.~E.,  1989, ApJ, 344, L53

\bibitem[\protect\citeauthoryear{{Peebles}}{{Peebles}}{1990}]{P90}
{Peebles} P.~J.~E.,  1990, ApJ, 362, 1

\bibitem[\protect\citeauthoryear{Pogorelov}{Pogorelov}{1978}]{P78}
Pogorelov A.~V.,  1978, The {M}inkowski multidimensional problem.
V. H. Winston \& Sons, Washington, D.C.

\bibitem[\protect\citeauthoryear{Rachev}{Rachev}{1984}]{R84}
Rachev S.~T.,  1984, Teor. Veroyatnost. i Primenen., 29, 625

\bibitem[\protect\citeauthoryear{{Rockafellar}}{{Rockafellar}}{1970}]{R70}
{Rockafellar} R.~T.,  1970, {Convex Analysis}.
Princeton Univ. Press, Princeton, NJ

\bibitem[\protect\citeauthoryear{{Sathyaprakash}, {Sahni}, {Shandarin} \&
  {Fisher}}{{Sathyaprakash} et~al.}{1998}]{SSSF98}
{Sathyaprakash} B.~S.,  {Sahni} V.,  {Shandarin} S.,    {Fisher} K.~B.,  1998,
  ApJ, 507, L109

\bibitem[\protect\citeauthoryear{{Shandarin} \& {Sathyaprakash}}{{Shandarin} \&
  {Sathyaprakash}}{1996}]{SS96}
{Shandarin} S.~F.,  {Sathyaprakash} B.~S.,  1996, ASTRJ2, 467, L25

\bibitem[\protect\citeauthoryear{{Shandarin} \& {Zel'dovich}}{{Shandarin} \&
  {Zel'dovich}}{1989}]{SZ89}
{Shandarin} S.~F.,  {Zel'dovich} Y.~B.,  1989, Rev. Modern Phys., 61, 185

\bibitem[\protect\citeauthoryear{{Spergel}, {Verde}, {Peiris}, {Komatsu},
  {Nolta}, {Bennett}, {Halpern}, {Hinshaw}, {Jarosik}, {Kogut}, {Limon},
  {Meyer}, {Page}, {Tucker}, {Weiland}, {Wollack} \& {Wright}}{{Spergel}
  et~al.}{2003}]{Setal03}
{Spergel} D.~N.,  {Verde} L.,  {Peiris} H.~V.,  {Komatsu} E.,  {Nolta} M.~R.,
  {Bennett} C.~L.,  {Halpern} M.,  {Hinshaw} G.,  {Jarosik} N.,  {Kogut} A.,
  {Limon} M.,  {Meyer} S.~S.,  {Page} L.,  {Tucker} G.~S.,  {Weiland} J.~L.,
  {Wollack} E.,    {Wright} E.~L., 2003, First year Wilkinson microwave
  anisotropy probe (WMAP) observations: determination of cosmological
  parameters, preprint (astro-ph/0302209)

\bibitem[\protect\citeauthoryear{{Susperregi} \& {Binney}}{{Susperregi} \&
  {Binney}}{1994}]{SB94}
{Susperregi} M.,  {Binney} J.,  1994, MNRAS, 271, 719

\bibitem[\protect\citeauthoryear{{Taylor} \& {Valentine}}{{Taylor} \&
  {Valentine}}{1999}]{TV99}
{Taylor} A.,  {Valentine} H.,  1999, MNRAS, 306, 491

\bibitem[\protect\citeauthoryear{Tomizawa}{Tomizawa}{1971}]{T71}
Tomizawa N.,  1971, Networks, 1, 173

\bibitem[\protect\citeauthoryear{{Tully}}{{Tully}}{1988}]{T88}
{Tully} R.~B.,  1988, {Nearby galaxies catalog}.
Cambridge University Press, Cambridge and New York

\bibitem[\protect\citeauthoryear{{Valentine}, {Saunders} \&
  {Taylor}}{{Valentine} et~al.}{2000}]{VST00b}
{Valentine} H.,  {Saunders} W.,    {Taylor} A.,  2000, MNRAS, 319, L13

\bibitem[\protect\citeauthoryear{{Vergassola}, {Dubrulle}, {Frisch} \&
  {Noullez}}{{Vergassola} et~al.}{1994}]{VDFN94}
{Vergassola} M.,  {Dubrulle} B.,  {Frisch} U.,    {Noullez} A.,  1994, A\&A,
  289, 325

\bibitem[\protect\citeauthoryear{Weber}{Weber}{1900}]{EMW00}
Weber E.~v.,  1900, in Burkhardt H.,  ed., Encyklop{\"a}die der Mathematischen
  Wissenschaften, Vol.~2, Analysis.
B.G.~Teubner, Leipzig, p.~294

\bibitem[\protect\citeauthoryear{{Weinberg}}{{Weinberg}}{1992}]{W92}
{Weinberg} D.~H.,  1992, MNRAS, 254, 315

\bibitem[\protect\citeauthoryear{{Weinberg} \& {Gunn}}{{Weinberg} \&
  {Gunn}}{1990}]{WG90}
{Weinberg} D.~H.,  {Gunn} J.~E.,  1990, MNRAS, 247, 260

\bibitem[\protect\citeauthoryear{{Whiting}}{{Whiting}}{2000}]{W00}
{Whiting} A.~B.,  2000, ApJ, 533, 50

\bibitem[\protect\citeauthoryear{{Zel'dovich}}{{Zel'dovich}}{1970}]{Z70}
{Zel'dovich} Y.~B.,  1970, A\&A, 5, 84

\end{thebibliography}

\end{document}